\documentclass[11pt,a4paper]{article}

\usepackage{jheppub}
\usepackage[utf8]{inputenc}
\usepackage[english]{babel}
\usepackage{mathrsfs}
\usepackage{slashed}
\usepackage{verbatim}
\usepackage{multirow}
\usepackage{braket}

\newcommand{\bm}[1]{\mbox{\boldmath $#1$}}
\newcommand{\st}{{\scriptscriptstyle T}}
\def\nn{\nonumber}
\def\cd{{\cdot}}
\DeclareMathOperator{\tr}{Tr}

\usepackage[normalem]{ulem} 
\renewcommand\sout{\bgroup \color[rgb]{0.55,0.00,0.99} \ULdepth=-.5ex \ULset}

\allowdisplaybreaks

\begin{document}

\preprint{NIKHEF 2016-030}

\title{Gluon and Wilson loop TMDs for hadrons of spin $\leq$ 1}

\author[a]{Dani\"el Boer,}
\author[b,c]{Sabrina Cotogno,}
\author[b,c]{Tom van Daal,}
\author[b,c]{Piet J. Mulders,}
\author[b,c]{\\Andrea Signori,}
\author[b,c,d]{and Ya-Jin Zhou}

\affiliation[a]{Van Swinderen Institute for Particle Physics and Gravity, University of Groningen, Nijenborgh 4, NL-9747 AG Groningen, The Netherlands}
\affiliation[b]{Department of Physics and Astronomy, VU University Amsterdam, De Boelelaan 1081, NL-1081 HV Amsterdam, The Netherlands}
\affiliation[c]{Nikhef, Science Park 105, NL-1098 XG Amsterdam, The Netherlands}
\affiliation[d]{School of Physics \& Key Laboratory of Particle Physics and Particle Irradiation (MOE), Shandong University, Jinan, Shandong 250100, China}

\emailAdd{d.boer@rug.nl}
\emailAdd{scotogno@nikhef.nl}
\emailAdd{tvdaal@nikhef.nl}
\emailAdd{mulders@few.vu.nl}
\emailAdd{asignori@nikhef.nl}
\emailAdd{zhouyj@sdu.edu.cn}

\abstract{In this paper we consider the parametrizations of gluon transverse momentum dependent (TMD) correlators in terms of TMD parton distribution functions (PDFs). These functions, referred to as TMDs, are defined as the Fourier transforms of hadronic matrix elements of nonlocal combinations of gluon fields. The nonlocality is bridged by gauge links, which have characteristic paths (future or past pointing), giving rise to a process dependence that breaks universality. For gluons, the specific correlator with one future and one past pointing gauge link is, in the limit of small $x$, related to a correlator of a single Wilson loop. We present the parametrization of Wilson loop correlators in terms of Wilson loop TMDs and discuss the relation between these functions and the small-$x$ `dipole' gluon TMDs. This analysis shows which gluon TMDs are leading or suppressed in the small-$x$ limit. We discuss hadronic targets that are unpolarized, vector polarized (relevant for spin-$1/2$ and spin-$1$ hadrons), and tensor polarized (relevant for spin-$1$ hadrons). The latter are of interest for studies with a future Electron-Ion Collider with polarized deuterons.}

\keywords{QCD Phenomenology, Spin and Polarization Effects, Wilson loops}

\maketitle

\section{Introduction}
In high energy collisions gluons become more important with increasing energy, due to the decreasing longitudinal momentum fraction $x$ that is typically being probed. This region has for example been studied by experiments at the Hadron-Electron Ring Accelerator (HERA) in inclusive deep inelastic scattering (DIS) and currently by experiments at the Large Hadron Collider (LHC) in proton-proton collisions. In less inclusive processes one can in addition become sensitive to the transverse momentum distribution of gluons. There is a rich variety of gluon transverse momentum dependent (TMD) parton distribution functions (PDFs), or TMDs for short, especially if one includes the polarization of hadrons. At the Relativistic Heavy Ion Collider (RHIC), experiments with spin-polarized protons are conducted and in future experiments, such as at an Electron-Ion Collider (EIC), polarized deuteron beams may be used.  For this reason it is useful to parametrize gluon TMD correlators as efficiently and systematically as possible for unpolarized, vector, and tensor polarized hadrons and to consider specifically the small-$x$ region. This is the intention of this paper.

In the present work, the starting point for the gluon TMD correlators are Fourier transforms of hadronic matrix elements of field strength tensors connected by Wilson lines or gauge links~\cite{Belitsky:2002sm,Boer:2003cm,Efremov:1978cu,Collins:1982wa,Boer:1999si,Collins:2002kn} that bridge the nonlocality of the field operators, ensuring color gauge invariance. The nonlocality includes transverse directions~\cite{Belitsky:2002sm,Boer:2003cm}, in which case one can consider, besides gluon-gluon correlators, also the matrix element of a single Wilson loop operator, which in this work is referred to as the Wilson loop correlator. The gauge invariant correlators are parametrized in terms of TMDs, depending on the longitudinal momentum fraction $x$ and the transverse momentum $k_\st^2$~\cite{Mulders:2000sh}. Including transverse momentum dependence, i.e.\ going beyond collinear kinematics, gives rise to a wealth of azimuthal asymmetries. This is particularly true when polarization degrees of freedom of the hadrons involved are considered, giving for instance rise to single spin asymmetries~\cite{Brodsky:2002cx,Bacchetta:2005rm,Collins:2002kn,Boer:1997nt,Boer:2003tx}. The parametrizations in terms of TMDs have been extensively studied, especially for the quark case, for different polarizations of hadrons up to and including spin $1$~\cite{Mulders:1995dh,Boer:1997nt,Ralston:1979ys,Sivers:1989cc,Collins:1992kk,Kotzinian:1994dv,Tangerman:1994eh,Bacchetta:2000jk}. In the collinear case, the parametrization in terms of PDFs for gluons in tensor polarized spin-1 hadrons has first been considered in  refs.~\cite{Jaffe:1989xy,Artru:1989zv}. A further proliferation of TMDs comes from the structure, i.e.\ the path dependence, of the gauge links. The gauge links depend on the process and as a consequence they give rise to observable process dependence and thus to a proliferation of TMDs. Since the dependence can be traced to the color flow in the hard scattering process, it is in principle possible to unravel this dependence~\cite{Collins:2002kn,Bomhof:2006dp,Buffing:2012sz,Buffing:2013kca}. In some cases one may find how different TMDs and processes are related, but in some cases TMDs with different gauge links are not related at all, encoding independent information~\cite{Boer:2016fqd}.

Here we limit ourselves to TMDs appearing in those contributions to the cross sections that are leading in inverse powers of the hard scale, referred to as leading twist TMDs. We will not be concerned with higher twist contributions~\cite{Jaffe:1996zw} nor with QCD corrections that are of higher order in the strong coupling $\alpha_s$, relevant for the evolution and the large transverse momentum region~\cite{Collins:2011zzd,Echevarria:2015uaa,Cherednikov:2012yd}. The higher twist contributions would generally involve correlators with more fields. In order to facilitate the study of the evolution of these TMDs we will discuss the transition to impact parameter space, without further studying the evolution itself. We present the parametrizations of the gluon-gluon and Wilson loop TMD correlators in terms of TMDs of definite rank for unpolarized, vector polarized, and tensor polarized hadrons, the latter being considered here for the first time. We also provide a new treatment of the connection between the gluon-gluon correlator at small $x$ and the Wilson loop correlator. This confirms the results of some specific examples that have been discussed in an earlier paper~\cite{Boer:2015pni}.

\section{Parametrizations of gluonic TMD correlators}
In 2001, Mulders and Rodrigues~\cite{Mulders:2000sh} presented the first parametrization of the gluon-gluon light-front correlator in terms of TMDs considering both unpolarized and vector polarized hadrons. In 2007, a different nomenclature for those TMDs was proposed by Mei{\ss}ner, Metz, and Goeke in ref.~\cite{Meissner:2007rx}, in close analogy to the ones for quarks. In this section we extend the analyses of refs.~\cite{Mulders:2000sh,Meissner:2007rx} by parametrizing both the gluon-gluon and Wilson loop correlators for unpolarized, vector polarized, as well as tensor polarized hadrons. The light-front correlators are expanded in a Lorentz basis of completely symmetric traceless tensors built from the partonic momentum $k_\st$ (see appendix~\ref{ss:STT} for the definitions of the relevant symmetric traceless tensors), and are expressed in terms of TMDs. Furthermore, a more systematic way of naming the various TMDs is introduced, keeping and extending the notation proposed in ref.~\cite{Meissner:2007rx}.

We start with outlining the most relevant variables. We denote by $P$ and $k$ the hadron and parton momenta respectively. 
We parametrize $k$ in terms of the dimensionful vectors $P$ and $n$, where $n$ is a lightlike vector satisfying $n^2 = 0$ and $P \cd n = 1$:
\begin{equation}
    k^\mu = x P^\mu + k_\st^\mu + (k\cd P - xM^2) \,n^\mu, 
\end{equation}
where $M$ is the mass of the hadron. The transverse direction is projected out using the metric tensor in transverse space, $g_\st^{\mu\nu} \equiv g^{\mu\nu} - P^{\{\mu} n^{\nu\}}$ (curly brackets denote symmetrization of the indices), with nonvanishing elements $g_\st^{11} = g_\st^{22} = -1$. For a polarized hadron we employ a spin vector $S$ needed to describe vector polarization for any hadron with spin $\ge 1/2$ and a symmetric traceless spin tensor $T$ to describe tensor polarization for hadrons with spin $\ge 1$~\cite{Bacchetta:2000jk,Bacchetta:2002}. We again parametrize $S$ and $T$ in terms of the dimensionful vectors $P$ and $n$,\footnote{We use the definition of $S_{LL}$ that is used in ref.~\cite{Bacchetta:2000jk}, which differs by a numerical factor from the definition in ref.~\cite{Bacchetta:2002}.}
\begin{align}
    S^\mu &= S_L \frac{P^\mu}{M} + S_T^\mu - MS_L \,n^\mu, \\
    T^{\mu\nu} &= \frac{1}{2} \left[ \frac{2}{3} S_{LL} \,g_\st^{\mu\nu} + \frac{4}{3} S_{LL} \frac{P^\mu P^\nu}{M^2} + \frac{S_{LT}^{\{\mu}P^{\nu\}}}{M} + S_{TT}^{\mu\nu} \right. \nn \\
    &\left. \qquad\;\; - \,\frac{4}{3} S_{LL} P^{\{\mu}n^{\nu\}} - M S_{LT}^{\{\mu}n^{\nu\}} + \frac{4}{3} M^2 S_{LL} \,n^\mu n^\nu \vphantom{\frac{P^\mu P^\nu}{M^2}} \right] ,
\end{align}
ensuring the relations
\begin{equation}
    P^2 = M^2 , \quad P \cd S = 0 , \quad P_\mu T^{\mu\nu} = 0 .
\end{equation}
For a spin-$1/2$ hadron only a spin vector is needed to parametrize the density matrix. For a spin-$1$ hadron also a tensor is required. While the spin vector $S$ for a spin-$1$ hadron signals a polarized hadron with $m = 1$ along that direction (in case of its length being one), the spin tensor $T$ corresponds to particular combinations of spin states (see e.g.\ refs.~\cite{Bacchetta:2000jk,Bacchetta:2002}). The spin tensor has five independent parameters, namely $S_{LL}$, the two components of the transverse vector $S_{LT}$, and the two independent components of the symmetric traceless transverse tensor $S_{TT}$.

We note that one could reinstate the combination $P\cd n = P^+$ by replacing everywhere $n \to n/P\cd n$. Introducing $\bar n \equiv (P-\tfrac{1}{2}M^2\,n)/P\cd n$, such that $n\cd\bar n = 1$, one can work with light cone components $a^+ = a\cd n$ and $a^- = a\cd\bar n$. Hence, in the infinite momentum frame $\bar n$ corresponds to the target hadron direction and $n$ to the conjugate direction. They are defined frame independently, however. In order to get the more natural interpretation in the hadron rest frame, one has the covariantly defined time- and spacelike directions,
\begin{equation}
    \hat t \equiv \frac{P}{M}, \quad \hat z \equiv \frac{P}{M} - M\,n,
\end{equation}
which become the standard time and spatial $z$-directions in the hadron rest frame. They are useful since the spin vector and tensor only contain the spacelike combination $\hat z$:
\begin{align}
    S^\mu &= S_L \,\hat z^\mu + S_T^\mu , \nn \\
    T^{\mu\nu} &= \frac{1}{2} \left[\frac{4}{3} \,S_{LL}\left(\hat z^\mu\hat z^\nu + \frac{1}{2} \,g_\st^{\mu\nu}\right) + \hat z^{\{\mu}S_{LT}^{\nu\}} + S_{TT}^{\mu\nu} \right] .
\end{align}

\subsection{Unpolarized hadrons}
\subsubsection{The gluon-gluon correlator}
For a color gauge invariant description of gluon correlations in hadrons one can consider the (unintegrated) gluon-gluon correlator as a starting point,
\begin{equation}
    \Gamma^{[U,U^\prime]\,\mu\nu;\rho\sigma}(k;P,n) \equiv \int \frac{d^4\xi}{(2\pi)^4} \,e^{ik\cdot\xi} \bra{P} F^{\mu\nu}(0) U_{[0,\xi]}^{\phantom{\prime}} F^{\rho\sigma}(\xi) U_{[\xi,0]}^\prime \ket{P} ,
\label{e:gamma_up}
\end{equation}
where color summation, a trace in color space (${\rm Tr}_c$), is implicitly assumed. The Wilson lines $U_{[0,\xi]}^{\phantom{\prime}}$ and $U_{[\xi,0]}^\prime$ guarantee color gauge invariance. Even though without specifying a process the path integrations could run along arbitrary paths, we have already included a dependence on the lightlike four-vector $n$, that enters upon consideration of staple-like gauge links running along the light-front ($\xi\cd n = 0$) via lightlike $\xi\cd P = \pm \infty$. A possible parametrization of the unintegrated correlator in eq.~\eqref{e:gamma_up}, constrained by hermiticity and parity conservation and respecting relations induced by time reversal (see appendix~\ref{a:constraints}), is\footnote{Relevant mass dimensions are $[\Gamma] = -2$ and $[A_i] = -4$.}
\begin{align}
    \Gamma^{[U,U^\prime]\,\mu\nu;\rho\sigma}(k;P,n) =& \;M^2 A_1 \,\epsilon^{\mu\nu\alpha\beta} \epsilon_{\phantom{\rho\sigma}\alpha\beta}^{\rho\sigma} + A_2 \,P^{[\mu} g^{\nu][\rho} P^{\sigma]} + A_3 \,k^{[\mu} g^{\nu][\rho} k^{\sigma]} \nn \\
    &+ (A_4 + iA_5) \,P^{[\mu} g^{\nu][\rho} k^{\sigma]} + (A_4 - iA_5) \,k^{[\mu} g^{\nu][\rho} P^{\sigma]} \nn \\
    &+ (A_6/M^2) \,P^{[\mu} k^{\nu]} P^{[\rho} k^{\sigma]} + M^4 A'_7 \,n^{[\mu} g^{\nu][\rho} n^{\sigma]} \nn \\
    &+ M^2 (A'_8 + iA'_9) \,P^{[\mu} g^{\nu][\rho} n^{\sigma]} + M^2 (A'_8 - iA'_9) \,n^{[\mu} g^{\nu][\rho} P^{\sigma]} \nn \\
    &+ M^2 (A'_{10} + iA'_{11}) \,k^{[\mu} g^{\nu][\rho} n^{\sigma]} + M^2 (A'_{10} - iA'_{11}) \,n^{[\mu} g^{\nu][\rho} k^{\sigma]} \nn \\
    &+ M^2 A'_{12} \,P^{[\mu} n^{\nu]} P^{[\rho} n^{\sigma]} + M^2 A'_{13} \,k^{[\mu} n^{\nu]} k^{[\rho} n^{\sigma]} \nn \\
    &+ (A'_{14} + iA'_{15}) \,P^{[\mu} k^{\nu]} P^{[\rho} n^{\sigma]} + (A'_{14} - iA'_{15}) \,P^{[\mu} n^{\nu]} P^{[\rho} k^{\sigma]} \nn \\
    &+ (A'_{16} + iA'_{17}) \,P^{[\mu} k^{\nu]} k^{[\rho} n^{\sigma]} + (A'_{16} - iA'_{17}) \,k^{[\mu} n^{\nu]} P^{[\rho} k^{\sigma]} \nn \\   
    &+ M^2 (A'_{18} + iA'_{19}) \,P^{[\mu} n^{\nu]} k^{[\rho} n^{\sigma]} \nn \\  
    &+ M^2 (A'_{18} - iA'_{19}) \,k^{[\mu} n^{\nu]} P^{[\rho} n^{\sigma]} ,  
    \label{e:par_up}
\end{align}
where $A_i = A_i(k\cd n, k\cd P, k^2)$ and the completely antisymmetric Levi-Civita tensor $\epsilon^{\mu\nu\rho\sigma}$ is fixed by taking $\epsilon^{-+12} = 1$. Terms with coefficients $A_5, A'_9, A'_{11}, A'_{15}, A'_{17}, A'_{19}$ are $T$-odd, and a prime on the coefficient indicates that the corresponding Lorentz structure includes the four-vector $n$. As it turns out, these structures do not give rise to any leading twist TMDs (see ref.~\cite{Goeke:2003az} for the analogous case for quarks). As we are only interested in leading twist functions, we will later on omit the terms containing $n$ from our description of the gluon-gluon correlators in case of polarized hadrons.

Integrating eq.~\eqref{e:gamma_up} over $k\cd P$, one obtains the TMD (light-front) correlator
\begin{equation}
    \Gamma^{[U,U^\prime]\,\mu\nu;\rho\sigma}(x,\bm{k}_\st;P,n) \equiv \int \left. \frac{d\xi\cd P\, d^2\xi_\st}{(2\pi)^3} \,e^{ik\cdot\xi} \bra{P} F^{\mu\nu}(0) U_{[0,\xi]}^{\phantom{\prime}} F^{\rho\sigma}(\xi) U_{[\xi,0]}^\prime \ket{P} \right|_{\xi\cd n=0} .
\end{equation}
The relevant correlator showing up in leading terms in the inverse hard scale can be recognized by counting $P\propto Q$ and $n \propto 1/Q$, with $Q$ denoting the hard scale. Suppressing the $P$ and $n$ dependence, which of course is present in the definition of transverse directions and in the paths of the gauge links, the leading (usually referred to as leading twist) correlator is then
\begin{equation}
    \Gamma^{ij}(x,\bm{k}_\st) \equiv \Gamma^{[U,U^\prime]\,ni;nj}(x,\bm{k}_\st;P,n) .
    \label{leadingcorrelator}
\end{equation}
Employing constant or symmetric traceless tensors, the light-front correlator is parametrized in terms of leading twist (i.e.\ twist-2) TMDs \emph{of definite rank}. For the unpolarized correlator one obtains
\begin{equation}
    \Gamma^{ij}(x,\bm{k}_\st) = \frac{x}{2} \left[ - \,g_\st^{ij} \,f_1(x,\bm{k}_\st^2) + \frac{k_\st^{ij}}{M^2} \,h_1^{\perp}(x,\bm{k}_\st^2) \right] ,
    \label{e:gamma_up_tmds}
\end{equation}
where the expressions of the TMDs in terms of the coefficients $A_i$ can be found in appendix~\ref{a:distributions}. Throughout this paper, the remaining dependence of TMDs on the gauge link as well as a reference to gluons, such as in $f_1^{g[U,U']}(x,\bm{k}_\st^2)$, is implicitly assumed, so we often simply write $f_1(x,\bm{k}_\st^2)$, etc.

We note that integration over $k_\st$ in eq.~\eqref{e:gamma_up_tmds} leads to the collinear correlator
\begin{equation}
    \Gamma^{ij}(x) \equiv \int d^2k_\st \,\Gamma^{ij}(x,\bm{k}_\st) = -\frac{x g_\st^{ij}}{2} \,f_1(x) ,
\end{equation}
parametrized in terms of a collinear PDF. Integrating over $k\cd n = x$ shows that this normalization is in agreement with the momentum sum rule for gluons taking the form 
\begin{equation}
    0 \le \int_0^1 dx \,x f_1(x) \le 1 ,
\end{equation}
which is not saturated because there is also a contribution from quarks.

\subsubsection{The Wilson loop correlator} \label{s:wilsonloop_up}
Again we start with a fully unintegrated correlator, now containing a Wilson loop operator,
\begin{equation}
    \Gamma_0^{[{\rm loop}]}(k;P) \equiv \int \frac{d^4\xi}{(2\pi)^4} \,e^{ik\cdot\xi} \bra{P} U^{[{\rm loop}]}  \ket{P} ,
    \label{e:wilsonloop-unint}
\end{equation}
where we implicitly include color tracing. The above quantity is a path-dependent quantity that reduces to the normalization $N_c \langle P\vert P\rangle$ upon integration over $d^4k$. In certain processes the latter contribution is subtracted, involving the operator $U^{[{\rm loop}]} - I$, such as in diffractive scattering \cite{Hebecker:1999ej}. As this subtraction only matters at $k=0$, we will not consider it here. To make contact with TMD correlators we can construct the loop from two staple-like paths along $n$, possibly including additional (color averaged) loops~\cite{Buffing:2013kca} in $U_{[0,\xi]}^{\phantom{\prime}} U_{[\xi,0]}^\prime$, but now without `parton' fields residing at $0$ and $\xi$,
\begin{equation}
    \Gamma_0^{[U,U^\prime]}(k;P,n) \equiv \int \frac{d^4\xi}{(2\pi)^4} \,e^{ik\cdot\xi} \bra{P} U_{[0,\xi]}^{\phantom{\prime}} U_{[\xi,0]}^\prime \ket{P} .
    \label{e:gamma0-unint}
\end{equation}
In the unintegrated amplitude expansion constrained by hermiticity and parity conservation for an unpolarized hadron, just one ($T$-even) amplitude remains:\footnote{Relevant mass dimensions are $[\Gamma_0] = -6$ and $[B_i] = -4$.}
\begin{equation}
    \Gamma_0^{[U,U^\prime]}(k,P,n) = \frac{B_1}{M^2} ,
\end{equation}
with $B_1 = B_1(k\cd n,k\cd P,k^2)$. The absence of the `parton' fields and the structure of the loop on the light-front still allows integration over $k\cd P$, and invariance in the $\xi\cd P$ direction implies a delta function $\delta(k\cd n)$:
\begin{align}
    \Gamma_0^{[U,U^\prime]}(x,\bm{k}_\st;P,n) &\equiv \left. \int \frac{d\xi\cd P\,d^2\xi_\st}{(2\pi)^3} \,e^{ik\cdot\xi} \bra{P} U_{[0,\xi]}^{\phantom{\prime}} U_{[\xi,0]}^\prime \ket{P}\right|_{\xi\cd n = 0} \nn \\
    &= \delta(x)\,\Gamma_0^{[U,U^\prime]}(\bm k_\st;P,n) ,
    \label{e:gamma0_up}
\end{align}
where the loop correlator integrated over $k\cd P$ and $k\cd n$ is given by
\begin{equation}
    \Gamma_0^{[U,U^\prime]}(\bm k_\st;P,n) \equiv \left. \int \frac{d^2\xi_\st}{(2\pi)^2} \,e^{ik_\st\cdot\xi_\st} \bra{P} U_{[0,\xi]}^{\phantom{\prime}} U_{[\xi,0]}^\prime \ket{P}\right|_{\xi\cd n = 0} .
    \label{e:gamma0_kt}
\end{equation}
Note that this correlator allows for azimuthal dependence in $\bm{k}_\st$. In the limit $x \to 0$ we have $t \equiv k^2 = k_\st^2 = -\bm k_\st^2$ (see also appendix~\ref{a:distributions}). Bearing in mind the proportionality to the longitudinal extent $L$ of the loop, $L \equiv \int d\xi\cd P = 2\pi\,\delta(0)$, the light-front correlator in eq.~(\ref{e:gamma0_kt}) is parametrized in terms of TMDs as follows (we suppress now the dependence on $P$ and $n$):
\begin{equation}
    \Gamma_0^{[U,U^\prime]}(\bm{k}_\st) = \frac{\pi L}{M^2} \,e(\bm{k}_\st^2) ,
    \label{e:gamma0_up_tmds}
\end{equation}
where the expression of the function $e$ in terms of the coefficient $B_1$ can be found in appendix~\ref{a:distributions}. The correlator in eq.~\eqref{e:gamma0_kt} appears for instance in the dipole cross section $\propto \int d^2 r_\perp/(2\pi)^2 \,e^{-i k_\perp\cdot r_\perp} \langle \tr U(0) U^\dagger(r_\perp) \rangle/N_c$, where $U$ is a gauge link running along the light cone from $-\infty$ to $+\infty$, up to endpoints forming a loop (see e.g.\ ref.~\cite{Dominguez:2011wm}).

In section~\ref{s:smallx} we will elaborate on the link between the Wilson loop operator and the gluon-gluon correlator at zero longitudinal momentum (i.e.\ $x=0$). We will consider two specific gauge links, namely a future and a past pointing staple-like gauge link, the simplest ones denoted by $[+]$ and $[-]$ respectively. These two gauge links also make up the rectangular Wilson loop $U^{[\Box]} \equiv U_{[0,\xi]}^{[+]} U_{[\xi,0]}^{[-]}$, consisting of Wilson lines running from $-\infty$ to $\infty$ along the $n$ direction, at some transverse separation $\xi_\st$. This loop can be written as a `square' of the form $O(0) O^{\dagger}(\xi)$ for a specific nonlocal operator $O$:
\begin{align}
    U^{[\Box]} &= U^n_{[-\infty,0_\st;\infty,0_\st]}U^T_{[\infty,0_\st;\infty,\xi_\st]}U^n_{[\infty,\xi_\st;-\infty,\xi_\st]}U^T_{[-\infty,\xi_\st;\infty,0_\st]} \nn \\
    &= \left( U^T_{[-\infty,\infty_\st;-\infty,0_\st]}U^n_{[-\infty,0_\st;\infty,0_\st]}U^T_{[\infty,0_\st;\infty, \infty_\st]} \right) \nn \\
    & \quad\; \times \left( U^T_{[-\infty,\infty_\st;-\infty, \xi_\st]}U^n_{[-\infty,\xi_\st;\infty,\xi_\st]}U^T_{[\infty,\xi_\st;\infty,\infty_\st]}\right)^\dagger ,
    \label{e:ubox}
\end{align} 
which are just the ingredients in the dipole operator~\cite{Dominguez:2011wm} now including transverse pieces. From eq.~\eqref{e:ubox} it follows that $e^{[\Box]}(\bm k_\st^2)$ is positive definite. Similarly, also $f_1^{[+,-]}(x, \bm k_\st^2)$ is positive definite.

\subsection{Vector polarized hadrons}
\subsubsection{The gluon-gluon correlator}
Let us now consider vector polarized hadrons. Since here we are only interested in vector polarization (we already discussed the unpolarized case), we would like to single out those terms from the parametrization of the correlator that describe a vector polarized hadron (i.e.\ terms containing $S$). To that end, we define
\begin{equation}
    \Delta\Gamma^{\mu\nu;\rho\sigma}(k;P,S) \equiv \frac{1}{2} \left[ \Gamma^{\mu\nu;\rho\sigma}(k;P,S) - \Gamma^{\mu\nu;\rho\sigma}(k;P,-S) \right] .
\end{equation}
A possible parametrization of this unintegrated correlator that is constrained by hermiticity and parity conservation and respects relations induced by time reversal (see appendix~\ref{a:constraints}) is\footnote{As already mentioned in the previous subsection, we omit gauge links for gluon-gluon correlators in the case of polarized hadrons, hence there is no dependence on the four-vector $n$ (which is the case in eq.~\eqref{e:par_up}). Since gauge links will always be present, we will, however, still allow for $T$-odd terms in the parametrizations.}
\begin{align}
    \Delta\Gamma^{\mu\nu;\rho\sigma}(k;P,S) =& - 2MA_7 \,\epsilon^{\mu\nu\rho\sigma} k \cd S + iMA_8 \left( \epsilon^{\mu\nu P[\rho} S^{\sigma]} - \epsilon^{\rho\sigma P[\mu} S^{\nu]} \right) \nn \\
    &+ iMA_9 \left( \epsilon^{\mu\nu S[\rho} P^{\sigma]} - \epsilon^{\rho\sigma S[\mu} P^{\nu]} \right) + iMA_{10} \left( \epsilon^{\mu\nu k[\rho} S^{\sigma]} - \epsilon^{\rho\sigma k[\mu} S^{\nu]} \right) \nn \\
    &+ iMA_{11} \left( \epsilon^{\mu\nu S[\rho} k^{\sigma]} - \epsilon^{\rho\sigma S[\mu} k^{\nu]} \right) + \frac{iA_{12}}{M} \left( \epsilon^{\mu\nu P[\rho} P^{\sigma]} - \epsilon^{\rho\sigma P[\mu} P^{\nu]} \right) k \cd S \nn \\
    &+ \frac{iA_{13}}{M} \left( \epsilon^{\mu\nu k[\rho} k^{\sigma]} - \epsilon^{\rho\sigma k[\mu} k^{\nu]} \right) k \cd S + \frac{iA_{14}}{M} \left( \epsilon^{\mu\nu P[\rho} k^{\sigma]} - \epsilon^{\rho\sigma P[\mu} k^{\nu]} \right) k \cd S \nn \\
    &+ \frac{iA_{15}}{M} \left( \epsilon^{\mu\nu k[\rho} P^{\sigma]} - \epsilon^{\rho\sigma k[\mu} P^{\nu]} \right) k \cd S + \frac{A_{16} + iA_{17}}{M} \,\epsilon^{\mu\nu PS} k^{[\rho} P^{\sigma]} \nn \\  
    &+ \frac{A_{16} - iA_{17}}{M} \,\epsilon^{\rho\sigma PS} k^{[\mu} P^{\nu]} + \frac{A_{18} + iA_{19}}{M} \,\epsilon^{\mu\nu kS} k^{[\rho} P^{\sigma]} \nn \\
    &+ \frac{A_{18} - iA_{19}}{M} \,\epsilon^{\rho\sigma kS} k^{[\mu} P^{\nu]} + \frac{A_{20} + iA_{21}}{M} \,\epsilon^{\mu\nu kP} P^{[\rho} S^{\sigma]} \nn \\
    &+ \frac{A_{20} - iA_{21}}{M} \,\epsilon^{\rho\sigma kP} P^{[\mu} S^{\nu]} + \frac{A_{22} + iA_{23}}{M} \,\epsilon^{\mu\nu kP} k^{[\rho} S^{\sigma]} \nn \\
    &+ \frac{A_{22} - iA_{23}}{M} \,\epsilon^{\rho\sigma kP} k^{[\mu} S^{\nu]} + \frac{A_{24} + iA_{25}}{M^3} \,\epsilon^{\mu\nu kP} k^{[\rho} P^{\sigma]} k \cd S \nn \\
    &+ \frac{A_{24} - iA_{25}}{M^3} \,\epsilon^{\rho\sigma kP} k^{[\mu} P^{\nu]} k \cd S ,
\end{align}
where we have employed the notation $\epsilon^{abcd} \equiv \epsilon^{\mu\nu\rho\sigma} a_\mu b_\nu c_\rho d_\sigma$ and square brackets denote antisymmetrization of the indices. The terms with coefficients $A_7, A_{16}, A_{18}, A_{20}, A_{22}, A_{24}$ are $T$-odd, and we note that the ones with coefficients $A_8$ up to $A_{15}$ are slightly different from those in ref.~\cite{Mulders:2000sh}.

Employing symmetric traceless tensors in $k_\st$, the light-front correlator is parametrized in terms of leading twist (i.e.\ twist-2), definite rank TMDs as follows (in analogy to eq.\ (\ref{leadingcorrelator})):
\begin{equation}
    \Delta\Gamma^{ij}(x,\bm{k}_\st) = \Delta\Gamma_L^{ij}(x,\bm{k}_\st) + \Delta\Gamma_T^{ij}(x,\bm{k}_\st) ,
    \label{e:gamma_vp_tmds}
\end{equation}
where\footnote{Throughout the paper, momenta indicated in boldface are two-dimensional vectors on the transverse plane rather than four-vectors. We define $k_\st^\mu = [0,0,\bm{k}_\st]$ etc., so that e.g.\ $\bm{k}_\st \cd \bm{S}_\st = - k_\st \cd S_\st$.}
\begin{align}
    \Delta\Gamma_L^{ij}(x,\bm{k}_\st) &= \frac{x}{2} \left[ i \epsilon_\st^{ij} S_L \,g_1(x,\bm{k}_\st^2) + \frac{{\epsilon_\st^{\{i}}_\alpha k_\st^{j\}\alpha} S_L}{2M^2} \,h_{1L}^\perp(x,\bm{k}_\st^2) \right] , \\
    \Delta\Gamma_T^{ij}(x,\bm{k}_\st) &= \frac{x}{2} \left[ - \,\frac{g_\st^{ij} \epsilon_\st^{S_\st k_\st}}{M} \,f_{1T}^\perp(x,\bm{k}_\st^2) + \frac{i \epsilon_\st^{ij} \bm{k}_\st \cd \bm{S}_\st}{M} \,g_{1T}(x,\bm{k}_\st^2) \right. \nn \\
    &\qquad\quad\!\! \left. - \,\frac{\epsilon_\st^{k_\st\{i} S_\st^{j\}} + \epsilon_\st^{S_\st\{i} k_\st^{j\}}}{4M} \,h_1(x,\bm{k}_\st^2) - \frac{{\epsilon_\st^{\{i}}_\alpha k_\st^{j\}\alpha S_\st}}{2M^3} \,h_{1T}^\perp(x,\bm{k}_\st^2) \right] ,
\end{align}
where $\epsilon_\st^{\mu\nu} \equiv \epsilon^{Pn\mu\nu}$, with nonzero components $\epsilon_\st^{12} = -\epsilon_\st^{21} = 1$. The expressions of the TMDs in terms of the coefficients $A_i$ can be found in appendix~\ref{a:distributions}. The functions $h_{1L}^\perp$, $f_{1T}^\perp$, $h_1$, and $h_{1T}^\perp$ are $T$-odd. The only surviving collinear PDF is the rank-$0$ function $g_1$, where we have omitted the index `$L$' on $g_{1} \equiv g_{1L}$. Note that $h_1 \neq h_{1T}$. The function $h_1$ now corresponds to the function $-\Delta H_T$ in the originally proposed parametrization in ref.~\cite{Mulders:2000sh}. The link to the more traditional parametrization is found by using the identity
\begin{equation}
    {\epsilon_\st^{\{i}}_\alpha k_\st^{j\}\alpha\beta} {S_\st}_\beta = \epsilon_\st^{k_\st\{i} k_\st^{j\}} \bm{k}_\st \cd \bm{S}_\st + \frac{1}{4} \bm{k}_\st^2 \left( S_\st^{\{j} \epsilon_\st^{i\}k_\st} + k_\st^{\{j} \epsilon_\st^{i\}S_\st} \right) .
\end{equation}
We can now recast eq.~\eqref{e:gamma_vp_tmds} into the more traditional, quite compact form
\begin{align}
    \Delta\Gamma^{ij}(x,\bm{k}_\st) &= \frac{x}{2} \left[ \frac{g_\st^{ij} \,\epsilon_\st^{k_\st S_\st}}{M} \,f_{1T}^\perp(x,\bm{k}_\st^2) + i \epsilon_\st^{ij} \,g_{1s}(x,\bm{k}_\st^2) \right. \nn \\
    &\qquad\quad\!\! \left. - \,\frac{\epsilon_\st^{k_\st\{i} S_\st^{j\}} + \epsilon_\st^{S_\st\{i} k_\st^{j\}}}{4M}\,h_{1T}(x,\bm{k}_\st^2) - \frac{\epsilon_\st^{k_\st\{i} k_\st^{j\}}}{2M^2} \,h_{1s}^\perp(x,\bm{k}_\st^2) \right] ,
    \label{e:gamma_vp_tmds_old}
\end{align}
where we have made use of the shorthand notation
\begin{equation}
    g_{1s}(x,\bm k_\st^2) \equiv S_L \,g_{1L}(x,\bm k_\st^2) + \frac{\bm k_\st\cd\bm S_\st}{M} \,g_{1T}(x,\bm k_\st^2) ,
\end{equation}
and likewise for $h_{1s}^\perp$. The functions $h_1$ and $h_{1T}$ are related as
\begin{equation}
    h_1(x,\bm k_\st^2) \equiv h_{1T}(x,\bm k_\st^2) + \frac{\bm{k}_\st^2}{2M^2} \,h_{1T}^\perp(x,\bm k_\st^2) .
\end{equation}
The function $h_1$ is a rank-1 function, $h_{1T}$ contains both rank-1 and rank-3 pieces, and $h_{1T}^\perp$ is a rank-3 function. Note that the function $h_1$ for gluons is, in spite of similarity in name, quite different from the quark transverse polarization (transversity) function $h_1$. 

\subsubsection{The Wilson loop correlator}
For the same reason as in the case of the gluon-gluon correlator, we define
\begin{equation}
    \Delta\Gamma_0^{[U,U^\prime]}(k;P,S,n) \equiv \frac{1}{2} \left[ \Gamma_0^{[U,U^\prime]}(k;P,S,n) - \Gamma_0^{[U,U^\prime]}(k;P,-S,n) \right] .
\end{equation}
A possible parametrization of this unintegrated correlator that is constrained by hermiticity and parity conservation and respects relations induced by time reversal (see appendix~\ref{a:constraints}) is
\begin{equation}
    \Delta\Gamma_0^{[U,U^\prime]}(k;P,S,n) = \frac{B_2}{M^3} \,\epsilon^{nPkS} ,
\end{equation}
which is a $T$-odd term. 

The loop correlator integrated over $k\cd P$ and $k\cd n$ is parametrized in terms of TMDs as follows:
\begin{equation}
    \Delta\Gamma_0^{[U,U^\prime]}(\bm{k}_\st) = \frac{\pi L}{M^2} \,\frac{\epsilon_\st^{S_\st k_\st}}{M} \,e_T(\bm{k}_\st^2) ,
    \label{e:gamma0_vp_tmds}
\end{equation}
where the expression of the $T$-odd function $e_T$ in terms of the coefficient $B_2$ can be found in appendix~\ref{a:distributions}.

\subsection{Tensor polarized hadrons}
\subsubsection{The gluon-gluon correlator}
We now include tensor polarization, which is relevant for spin-$1$ hadrons. Similarly as for the vector polarized case, we define
\begin{equation}
    \Delta\Gamma^{\mu\nu;\rho\sigma}(k;P,T) \equiv \frac{1}{2} \left[ \Gamma^{\mu\nu;\rho\sigma}(k;P,T) - \Gamma^{\mu\nu;\rho\sigma}(k;P,-T) \right] ,
\end{equation}
where we have taken the vector polarization to be zero (i.e.\ $S=0$). A possible parametrization of this unintegrated correlator that is constrained by hermiticity and parity conservation and respects relations induced by time reversal (see appendix~\ref{a:constraints}) is
\begin{align}
    \Delta\Gamma^{\mu\nu;\rho\sigma}(k;P,T) =& \;A_{26} \,k^{[\mu} T^{\nu][\rho} k^{\sigma]} + A_{27} \,P^{[\mu} T^{\nu][\rho} P^{\sigma]} + (A_{28} + iA_{29}) \,k^{[\mu} T^{\nu][\rho} P^{\sigma]} \nn \\
    &+ (A_{28} - iA_{29}) \,P^{[\mu} T^{\nu][\rho} k^{\sigma]} + \frac{A_{30} + iA_{31}}{M^2} \,k_{\alpha} T^{\alpha[\mu} k^{\nu]} k^{[\rho} P^{\sigma]} \nn \\
    &+ \frac{A_{30} - iA_{31}}{M^2} \,k_{\alpha} T^{\alpha[\rho} k^{\sigma]} k^{[\mu} P^{\nu]} + \frac{A_{32} + iA_{33}}{M^2} \,k_{\alpha} T^{\alpha[\mu} P^{\nu]} k^{[\rho} P^{\sigma]} \nn \\
    &+ \frac{A_{32} - iA_{33}}{M^2} \,k_{\alpha} T^{\alpha[\rho} P^{\sigma]} k^{[\mu} P^{\nu]} + M^2 A_{34} \left( g^{\mu[\rho} T^{\sigma]\nu} - g^{\nu[\rho} T^{\sigma]\mu} \right) \nn \\
    &+ (A_{35} + iA_{36}) \,k_{\alpha} T^{\alpha[\mu} g^{\nu][\rho} k^{\sigma]} + (A_{35} - iA_{36}) \,k_{\alpha} T^{\alpha[\rho} g^{\sigma][\mu} k^{\nu]} \nn \\
    &+ (A_{37} + iA_{38}) \,k_{\alpha} T^{\alpha[\mu} g^{\nu][\rho} P^{\sigma]} + (A_{37} - iA_{38}) \,k_{\alpha} T^{\alpha[\rho} g^{\sigma][\mu} P^{\nu]} \nn \\
    &+ A_{39} \,k_{\alpha} k_{\beta} T^{\alpha\beta} \epsilon^{\mu\nu\kappa\lambda} \epsilon_{\phantom{\rho\sigma}\kappa\lambda}^{\rho\sigma} + \frac{A_{40}}{M^2} \,k_{\alpha} k_{\beta} T^{\alpha\beta} P^{[\mu} g^{\nu][\rho} P^{\sigma]} \nn \\
    &+ \frac{A_{41}}{M^2} \,k_{\alpha} k_{\beta} T^{\alpha\beta} k^{[\mu} g^{\nu][\rho} k^{\sigma]} + \frac{(A_{42} + iA_{43})}{M^2} \,k_{\alpha} k_{\beta} T^{\alpha\beta} P^{[\mu} g^{\nu][\rho} k^{\sigma]} \nn \\
    &+ \frac{(A_{42} - iA_{43})}{M^2} \,k_{\alpha} k_{\beta} T^{\alpha\beta} k^{[\mu} g^{\nu][\rho} P^{\sigma]} \nn \\
    &+ \frac{A_{44}}{M^4} \,k_{\alpha} k_{\beta} T^{\alpha\beta} P^{[\mu} k^{\nu]} P^{[\rho} k^{\sigma]} ,
\end{align}
where the terms with coefficients $A_{29},A_{31},A_{33},A_{36},A_{38},A_{43}$ are $T$-odd. 

The light-front correlator is parametrized in terms of leading twist (i.e.\ twist-2) TMDs of definite rank as follows:
\begin{equation}
    \Delta\Gamma^{ij}(x,\bm{k}_\st) = \Delta\Gamma_{LL}^{ij}(x,\bm{k}_\st) + \Delta\Gamma_{LT}^{ij}(x,\bm{k}_\st) + \Delta\Gamma_{TT}^{ij}(x,\bm{k}_\st) ,
    \label{e:gamma_tp_tmds}
\end{equation}
where
\begin{align}
    \Delta\Gamma_{LL}^{ij}(x,\bm{k}_\st) &= \frac{x}{2} \left[ - \,g_\st^{ij} S_{LL} \,f_{1LL}(x,\bm{k}_\st^2) + \frac{k_\st^{ij} S_{LL}}{M^2} \,h_{1LL}^{\perp}(x,\bm{k}_\st^2) \right] , \\
    \Delta\Gamma_{LT}^{ij}(x,\bm{k}_\st) &= \frac{x}{2} \left[ - \,\frac{g_\st^{ij} \bm{k}_\st \cd \bm{S}_{LT}}{M} \,f_{1LT}(x,\bm{k}_\st^2) 
    + \frac{i \epsilon_\st^{ij} \epsilon_\st^{S_{LT}k_\st}}{M} \,g_{1LT}(x,\bm{k}_\st^2) \right. \nn \\
    &\qquad\quad\!\! \left. + \,\frac{S_{LT}^{\{i} k_\st^{j\}}}{M} \,h_{1LT}(x,\bm{k}_\st^2) + \frac{k_\st^{ij\alpha} {S_{LT}}_\alpha}{M^3} \,h_{1LT}^{\perp}(x,\bm{k}_\st^2) \right] , \\
    \Delta\Gamma_{TT}^{ij}(x,\bm{k}_\st) &= \frac{x}{2} \left[ - \,\frac{g_\st^{ij} k_\st^{\alpha\beta} {S_{TT}}_{\alpha\beta}}{M^2} \,f_{1TT}(x,\bm{k}_\st^2) 
    + \frac{i \epsilon_\st^{ij} {\epsilon^{\beta}_\st}_\gamma k_\st^{\gamma\alpha} {S_{TT}}_{\alpha\beta}}{M^2} \,g_{1TT}(x,\bm{k}_\st^2) \right. \nn \\
    &\qquad\quad\!\! \left. + \,S_{TT}^{ij} \,h_{1TT}(x,\bm{k}_\st^2) + \frac{{S_{TT}^{\{i}}_\alpha k_\st^{j\}\alpha}}{M^2} \,h_{1TT}^{\perp}(x,\bm{k}_\st^2) \right. \nn \\
    &\qquad\quad\!\! \left. + \,\frac{k_\st^{ij\alpha\beta} {S_{TT}}_{\alpha\beta}}{M^4} \,h_{1TT}^{\perp\perp}(x,\bm{k}_\st^2) \right] .
\end{align}
The expressions of the TMDs in terms of the coefficients $A_i$ can be found in appendix~\ref{a:distributions}. The functions $g_{1LT}$ and $g_{1TT}$ are $T$-odd. In the collinear case the rank-$0$ functions $f_{1LL}$ and $h_{1TT}$ survive. The former function was also called $b_1$ in the quark case, and the latter function shows up in the structure function $\Delta(x,Q^2)$ discussed in ref.~\cite{Jaffe:1989xy} and is called $\Delta_2G(x)$ in ref.~\cite{Artru:1989zv}.

\subsubsection{The Wilson loop correlator}
Similarly to the vector polarized case, we define
\begin{equation}
    \Delta\Gamma_0^{[U,U^\prime]}(k;P,T,n) \equiv \frac{1}{2} \left[ \Gamma_0^{[U,U^\prime]}(k;P,T,n) - \Gamma_0^{[U,U^\prime]}(k;P,-T,n) \right] ,
\end{equation}
where we have taken the vector polarization to be zero (i.e.\ $S=0$). A possible parametrization of this unintegrated correlator that is constrained by hermiticity and parity conservation and respects relations induced by time reversal (see appendix~\ref{a:constraints}) is
\begin{equation}
    \Delta\Gamma_0^{[U,U^\prime]}(k;P,T,n) = \frac{B_3}{M^4} \,k_\mu k_\nu T^{\mu\nu} + B_4 \,n_\mu n_\nu T^{\mu\nu} + \frac{B_5}{M^2} \,k_\mu n_\nu T^{\mu\nu} .
\end{equation}

The loop correlator integrated over $k\cd P$ and $k\cd n$ is parametrized in terms of TMDs as follows:
\begin{equation}
    \Delta\Gamma_0^{[U,U^\prime]}(\bm{k}_\st) = \frac{\pi L}{M^2} \left[ S_{LL} \,e_{LL}(\bm{k}_\st^2) + \frac{\bm{k}_\st \cd \bm{S}_{LT}}{M} \,e_{LT}(\bm{k}_\st^2) + \frac{k_\st^{\alpha\beta} {S_{TT}}_{\alpha\beta}}{M^2} \,e_{TT}(\bm{k}_\st^2) \right] ,
    \label{e:gamma0_tp_tmds}
\end{equation}
where the expressions of the TMDs in terms of the coefficients $B_i$ can be found in appendix~\ref{a:distributions}.

\section{The gluon-gluon correlator at small \textit{x}} \label{s:smallx} 
In this section we discuss the relation between the gluon-gluon correlator at small $x$ and the Wilson loop correlator. This connection {\it only} applies to the gluon-gluon correlator with the staple-like $[+]$ and $[-]$ gauge links. In the Wilson loop correlator those gauge links constitute the rectangular Wilson loop $U^{[\Box]} \equiv U_{[0,\xi]}^{[+]} U_{[\xi,0]}^{[-]}$.

We will start from the Wilson loop correlator integrated over $k\cd P$ and $k\cd n$ given in eq.~\eqref{e:gamma0_kt}. To study its $k_\st$ dependence, we use the results in eq.~(15) of ref.~\cite{Buffing:2013kca} to calculate $k_\st^ik_\st^j\Gamma_0$. Performing one partial integration in $0$ and the other in $\xi$ and using the relevant gluonic pole factor $C_{GG}^{[\Box]} = 4$, we obtain
\begin{align}
    k_\st^i k_\st^j\,\Gamma_0^{[\Box]}(\bm{k}_\st) &= \left. 4 \int \frac{d^2\xi_\st}{(2\pi)^2} \,e^{ik_\st\cdot\xi_\st} \bra{P} G_T^i(0)\,U_{[0,\xi]}^{[+]}\,G_T^j(\xi)\,U_{[\xi,0]}^{[-]} \ket{P} \right|_{\xi\cd n = 0} \nonumber \\ 
    &= \int \left. \frac{d\eta\cd P\,d\eta^\prime\cd P\,d^2\xi_\st}{(2\pi)^2} \,e^{ik_\st\cdot\xi_\st} 
    \bra{P} F^{ni}(\eta^\prime)\,U^{[+]}_{[\eta^\prime,\eta]}\,F^{nj}(\eta)\,U^{[-]}_{[\eta,\eta^\prime]} \ket{P}\right|_{\hspace{-0.07cm} \begin{array}{l} \scriptstyle{\eta^\prime \cd n = \eta \cd n = 0,} \\ \scriptstyle{\eta^\prime_\st = 0_\st, \,\eta_\st = \xi_\st} \end{array}} \nonumber \\ 
    &= 2\pi L \int \left. \frac{d\xi\cd P\,d^2\xi_\st}{(2\pi)^3} \,e^{ik\cdot\xi} \bra{P} F^{ni}(0)\,U^{[+]}_{[0,\xi]}\,F^{nj}(\xi)\,U^{[-]}_{[\xi,0]} \ket{P}\right|_{\xi\cd n = k\cd n = 0} \nonumber \\ 
    &= 2\pi L \;\Gamma^{[+,-]\,ij}(0,\bm{k}_\st) ,
    \label{correlatorrel}
\end{align}
which implies that
\begin{equation}
    \Gamma^{[+,-]\,ij}(0,\bm k_\st) = \frac{k_\st^i k_\st^j}{2\pi L} \,\Gamma_0^{[\Box]}(\bm k_\st) .
\end{equation}
The dependence on $\bm k_\st$ is in fact in this limit just the dependence on $t \equiv k^2$, remaining after the integration over $k\cd n = x$ and $k\cd P$ (the mass spectrum of intermediate states). Thus it is appropriate to write the previous equation as
\begin{equation}
    \Gamma^{[+,-]\,ij}(x,\bm k_\st) \;\stackrel{x\to 0}{\longrightarrow}\; \frac{k_\st^i k_\st^j}{2\pi L} \left. \!\Gamma_0^{[\Box]}(\bm k_\st) \right|_{k_\st^2 = t} .
    \label{limit}
\end{equation}
The above results agree with the result in ref.~\cite{Dominguez:2011wm} where in the small-$x$ limit $f_1^{[+,-]}(x,\bm{k}_\st^2)$ becomes proportional to the dipole cross section. In ref.~\cite{Boer:2015pni} that connection was made on the correlator level for the case of a transversely polarized hadron, which corresponds to the above eq.~\eqref{correlatorrel} and will also be discussed below. 

For unpolarized hadrons the right-hand side of eq.~\eqref{limit} is given by the parametrization in eq.~\eqref{e:gamma0_up_tmds}. It follows that
\begin{eqnarray}
    \Gamma_U^{ij}(x,\bm{k}_\st) \;&=&\; \frac{x}{2} \left[ - \,g_\st^{ij} \,f_1(x,\bm{k}_\st^2) + \frac{k_\st^{ij}}{M^2} \,h_1^{\perp}(x,\bm{k}_\st^2) \right] \nn \\[5pt]
    \;&\stackrel{x\to 0}{\longrightarrow}&\; \frac{k_\st^i k_\st^j}{2M^2} \,e(\bm{k}_\st^2) \nn \\[5pt]
    \;&=&\; \frac{1}{2} \left[ - \,g_\st^{ij} \,\frac{\bm k_\st^2}{2M^2} \,e(\bm{k}_\st^2) + \frac{k_\st^{ij}}{M^2} \,e(\bm{k}_\st^2) \right] ,
\end{eqnarray}
which implies that
\begin{equation}
    \lim_{x\to0} \,x f_1(x,\bm{k}_\st^2) = \frac{\bm{k}_\st^2}{2M^2} \lim_{x\to0} \,x h_1^\perp(x,\bm{k}_\st^2) = \frac{\bm{k}_\st^2}{2M^2} \,e(\bm{k}_\st^2) .
    \label{e:upol_result}
\end{equation} 
This means that $h_1^\perp$ must be maximal \cite{Mulders:2000sh}, i.e.\ $h_1^\perp = 2 M^2 f_1/\bm{k}_\st^2$, as it is in fact the case in the small-$x$ $k_\st$-factorization approach \cite{Catani:1990eg} and in the framework of the color glass condensate \cite{Metz:2011wb}. This result indicates that the unpolarized dipole gluon distribution grows as $1/x$ towards small $x$, apart from subdominant modifications from resummation of large logarithms in $1/x$ and higher twist effects. 

For longitudinally polarized hadrons eq.~\eqref{limit} implies that $g_1$ and $h_{1L}^\perp$ are less divergent than $1/x$ in the limit of small $x$. For $g_1$ this is in accordance with the fact that in DGLAP and CCFM evolution the splitting kernel lacks the $1/x$ factor of the kernel of $f_1$, see e.g.\ ref.~\cite{Maul:2001uz}. Again this does not include resummation of large logarithms in $1/x$ leading to nonlinear evolution, which may alter the result in the very small $x$ region \cite{Bartels:1995iu,Bartels:1996wc,Kovchegov:2015pbl}.  
Now let us consider transversely polarized hadrons. The right-hand side of eq.~\eqref{limit} is given by the parametrization in eq.~\eqref{e:gamma0_vp_tmds}. We find for the symmetric part (symmetric in $i,j$) of the transversely polarized gluon-gluon correlator
\begin{eqnarray}
    {\Delta\Gamma_T^{ij}}_\text{sym}(x,\bm{k}_\st) \;&=&\; \frac{x}{2} \left[ - \,\frac{g_\st^{ij} \epsilon_\st^{S_\st k_\st}}{M} \,f_{1T}^\perp(x,\bm{k}_\st^2) - \frac{\epsilon_\st^{k_\st\{i} S_\st^{j\}} + \epsilon_\st^{S_\st\{i} k_\st^{j\}}}{4M} \,h_1(x,\bm{k}_\st^2) \right. \nn \\[5pt] 
    && \quad\;\;\, \left. - \,\frac{{\epsilon_\st^{\{i}}_\alpha k_\st^{j\}\alpha S_\st}}{2M^3} \,h_{1T}^\perp(x,\bm{k}_\st^2) \right] \nn \\[5pt]
    \;&\stackrel{x\to 0}{\longrightarrow}&\; \frac{k_\st^i k_\st^j}{2M^2} \,\frac{\epsilon_\st^{S_\st k_\st}}{M} \,e_T(\bm{k}_\st^2) \nn \\[5pt]
    \;&=&\; \frac{1}{2} \left[ - \,\frac{g_\st^{ij} \epsilon_\st^{S_\st k_\st}}{M} \,\frac{\bm{k}_\st^2}{2M^2} \,e_T(\bm{k}_\st^2) - \frac{\epsilon_\st^{k_\st\{i} S_\st^{j\}} + \epsilon_\st^{S_\st\{i} k_\st^{j\}}}{4M} \,\frac{\bm{k}_\st^2}{2M^2} \,e_T(\bm{k}_\st^2) \right. \nn \\[5pt]
    && \quad\;\;\, \left. + \,\frac{{\epsilon_\st^{\{i}}_\alpha k_\st^{j\}\alpha S_\st}}{2M^3} \,e_T(\bm{k}_\st^2) \right] ,
\end{eqnarray}
which implies that
\begin{align}
    \lim_{x\to0} \,x f_{1T}^\perp(x,\bm{k}_\st^2) &= \lim_{x\to0} \,x h_1(x,\bm{k}_\st^2) = - \frac{\bm{k}_\st^2}{2M^2} \lim_{x\to0} \,x h_{1T}^\perp(x,\bm{k}_\st^2) \nn \\
    &= \frac{1}{2} \lim_{x\to0} \,x h_{1T}(x,\bm k_\st^2) = \frac{\bm{k}_\st^2}{2M^2} \,e_T(\bm{k}_\st^2) ,
    \label{e:tpol_result}
\end{align}
in agreement with the leading logarithmic result of ref.~\cite{Boer:2015pni}. It involves the $C$-odd operator structure $U^{[\Box]} - U^{[\Box] \dag}$ (see appendix~\ref{a:constraints}) surviving in $\Delta \Gamma_0^{[\Box]}(k;P,S,n)$ for a transversely polarized proton, which is the dipole odderon operator~\cite{Hatta:2005as}. Therefore, this is also referred to as the spin-dependent odderon~\cite{Zhou:2013gsa}. The odderon operator $O_{1T}^\perp$ as defined in ref.~\cite{Boer:2015pni} and used in a model calculation in ref.~\cite{Szymanowski:2016mbq} is related to $e_T$ by $O_{1T}^\perp = \pi \,e_T / (2 M^2)$. The odderon in transverse spin asymmetries in elastic scattering has earlier been considered in refs.~\cite{Ryskin:1987ya,Buttimore:1998rj,Leader:1999ua}, but without discussion of its operator structure. As the only nonzero function in the unpolarized case is the even rank function $e(\bm{k}_\st^2)$, which survives for the $C$-even Wilson loop operator combination $U^{[\Box]} + U^{[\Box] \dag}$, it also follows that there appears no spin-independent odderon in this formalism, or rather that it is less divergent than $1/x$ in the limit of small $x$. This suggests that it will be suppressed in the small-$x$ limit compared to the $C$-even leading contribution. 

For spin-$1$ hadrons eq.~\eqref{limit} implies that at small $x$ three tensor polarized TMDs remain, while the rest becomes zero. To be specific, for longitudinal-longitudinal ($LL$) polarization we find
\begin{eqnarray}
    \Delta\Gamma_{LL}^{ij}(x,\bm{k}_\st) \;&=&\; \frac{x}{2} \left[ - \,g_\st^{ij} S_{LL} \,f_{1LL}(x,\bm{k}_\st^2) + \frac{k_\st^{ij} S_{LL}}{M^2} \,h_{1LL}^{\perp}(x,\bm{k}_\st^2) \right] \nn \\[5pt]
    \;&\stackrel{x\to 0}{\longrightarrow}&\;  \frac{k_\st^i k_\st^j}{2M^2} \,S_{LL} \,e_{LL}(\bm{k}_\st^2) \nn \\[5pt]
    \;&=&\; \frac{1}{2} \left[ - \,g_\st^{ij} S_{LL} \,\frac{\bm k_\st^2}{2M^2} \,e_{LL}(\bm{k}_\st^2) + \frac{k_\st^{ij} S_{LL}}{M^2} \,e_{LL}(\bm{k}_\st^2) \right],
\end{eqnarray}
which implies that
\begin{equation}
    \lim_{x\to0} \,x f_{1LL}(x,\bm{k}_\st^2) = \frac{\bm{k}_\st^2}{2M^2} \lim_{x\to0} \,x h_{1LL}^\perp(x,\bm{k}_\st^2) = \frac{\bm{k}_\st^2}{2M^2} \,e_{LL}(\bm{k}_\st^2).
    \label{e:LLpol_result}
\end{equation}

For the case of longitudinal-transverse ($LT$) polarization, we find for the symmetric part of the gluon-gluon correlator
\begin{eqnarray}
    {\Delta\Gamma_{LT}^{ij}}_\text{sym}(x,\bm{k}_\st) \;&=&\; \frac{x}{2} \left[ - \,\frac{g_\st^{ij} \bm{k}_\st \cd \bm{S}_{LT}}{M} \,f_{1LT}(x,\bm{k}_\st^2) + \frac{S_{LT}^{\{i} k_\st^{j\}}}{M} \,h_{1LT}(x,\bm{k}_\st^2) \right. \nn \\[5pt] 
    && \quad\;\;\, \left. + \,\frac{k_\st^{ij\alpha} {S_{LT}}_\alpha}{M^3} \,h_{1LT}^{\perp}(x,\bm{k}_\st^2) \right] \nn \\[5pt]
    \;&\stackrel{x\to 0}{\longrightarrow}&\; \frac{k_\st^i k_\st^j}{2M^2} \,\frac{\bm{k}_\st \cd \bm{S}_{LT}}{M} \,e_{LT}(\bm{k}_\st^2) \nn \\[5pt]
    \;&=&\; \frac{1}{2} \left[ - \,\frac{g_\st^{ij} \bm{k}_\st \cd \bm{S}_{LT}}{M} \,\frac{\bm{k}_\st^2}{4M^2} \,e_{LT}(\bm{k}_\st^2) + \frac{S_{LT}^{\{i} k_\st^{j\}}}{M} \,\frac{\bm{k}_\st^2}{4M^2} \,e_{LT}(\bm{k}_\st^2) \right. \nn \\[5pt] 
    && \quad\;\;\, \left. - \,\frac{k_\st^{ij\alpha} {S_{LT}}_\alpha}{M^3} \,e_{LT}(\bm{k}_\st^2) \right] ,
\end{eqnarray}
which implies that
\begin{equation}
    \lim_{x\to0} \,x f_{1LT}(x,\bm{k}_\st^2) = \lim_{x\to0} \,x h_{1LT}(x,\bm{k}_\st^2) = -\frac{\bm{k}_\st^2}{4M^2} \lim_{x\to0} \,x h_{1LT}^\perp(x,\bm{k}_\st^2) = \frac{\bm{k}_\st^2}{4M^2} \,e_{LT}(\bm{k}_\st^2) .
\end{equation}

For the case of transverse-transverse ($TT$) polarization, we find for the symmetric part of the gluon-gluon correlator
\begin{eqnarray}
    {\Delta\Gamma_{TT}^{ij}}_\text{sym}(x,\bm{k}_\st) \;&=&\; \frac{x}{2} \left[ - \,\frac{g_\st^{ij} k_\st^{\alpha\beta} {S_{TT}}_{\alpha\beta}}{M^2} \,f_{1TT}(x,\bm{k}_\st^2) + S_{TT}^{ij} \,h_{1TT}(x,\bm{k}_\st^2) \right. \nn \\[5pt] 
    && \quad\;\;\, \left. + \,\frac{{S_{TT}^{\{i}}_\alpha k_\st^{j\}\alpha}}{M^2} \,h_{1TT}^{\perp}(x,\bm{k}_\st^2) + \frac{k_\st^{ij\alpha\beta} {S_{TT}}_{\alpha\beta}}{M^4} \,h_{1TT}^{\perp\perp}(x,\bm{k}_\st^2) \right] \nn \\[5pt]
    \;&\stackrel{x\to 0}{\longrightarrow}&\; \frac{k_\st^i k_\st^j}{2M^2} \,\frac{k_\st^{\alpha\beta} {S_{TT}}_{\alpha\beta}}{M^2} \,e_{TT}(\bm{k}_\st^2) \nn \\[5pt]
    \;&=&\; \frac{1}{2} \left[ - \,\frac{g_\st^{ij} k_\st^{\alpha\beta} {S_{TT}}_{\alpha\beta}}{M^2} \,\frac{\bm{k}_\st^2}{6M^2} \,e_{TT}(\bm{k}_\st^2) + S_{TT}^{ij} \,\frac{\bm{k}_\st^4}{4M^4} \,e_{TT}(\bm{k}_\st^2) \right. \nn \\[5pt] 
    && \quad\;\;\, \left. - \,\frac{{S_{TT}^{\{i}}_\alpha k_\st^{j\}\alpha}}{M^2} \,\frac{\bm{k}_\st^2}{3M^2} \,e_{TT}(\bm{k}_\st^2) + \frac{k_\st^{ij\alpha\beta} {S_{TT}}_{\alpha\beta}}{M^4} \,e_{TT}(\bm{k}_\st^2) \right] ,
\end{eqnarray}
which implies that
\begin{align}
    \lim_{x\to0} \,x f_{1TT}(x,\bm{k}_\st^2) &= \frac{2M^2}{3\bm{k}_\st^2} \lim_{x\to0} \,x h_{1TT}(x,\bm{k}_\st^2) = - \frac{1}{2} \lim_{x\to0} \,x h_{1TT}^\perp(x,\bm{k}_\st^2) \nn \\
    &= \frac{\bm{k}_\st^2}{6M^2} \lim_{x\to0} \,x h_{1TT}^{\perp\perp}(x,\bm{k}_\st^2) = \frac{\bm{k}_\st^2}{6M^2} \,e_{TT}(\bm{k}_\st^2) .
\end{align}

\section{Summary and discussion} \label{s:summary}
We have parametrized the gluon light-front correlators in terms of definite rank TMDs using a basis of symmetric traceless tensors in $k_\st$. In table~\ref{t:summary} we list the leading twist TMDs (multiplied by $x$), their rank, and their behavior under time reversal and charge conjugation. The rank-$0$ functions are the ones that also appear as collinear PDFs. In the last column the $x\rightarrow 0$ limit is considered for the functions $xf_{\ldots}^{[+,-]}(x,\bm{k}_\st^2)$. We emphasize that the connection to the Wilson loop or dipole TMDs applies only to the TMDs with one future and one past pointing link. Some of these functions are expected to be zero at $x = 0$, others become equal to the TMDs $e_{\ldots}(\bm{k}_\st^2)$ in the Wilson loop operator. Conjecturing that the dependence on $\bm k_\st^2$ reflects the analytic behavior in $k^2$, it simplifies the picture for the gluon TMDs $xf_{\ldots}^{[+,-]}(x,\bm{k}_\st^2)$ at small $x$, several of them becoming proportional to one another. The $C$- and $T$-behavior of the TMDs in the gluon-gluon correlator and those in the Wilson loop correlator correctly match. The functions with a nonvanishing limit are expected to behave as $1/x$, or a slightly modified power after resummation of other small-$x$ effects, e.g.\ with an $\ln(1/x)$ behavior. The functions with a vanishing limit are the ones for longitudinally polarized hadrons as well as those linked to circular gluon polarization (the $g$-type TMDs). \\
\begin{table}[!htb]
\begin{center}
{\renewcommand{\arraystretch}{1.4}
\begin{tabular}{|c|c|c|c|c|c|c|c|}
\hline
& \textbf{Ref. \protect{\cite{Meissner:2007rx}}} & \textbf{Ref. \protect{\cite{Mulders:2000sh}}} & \textbf{Rank} & $\bm{T}$ & $\bm{C}$ & \textbf{Limit} $\bm{x \to 0}$ \\ \hline \hline
$xf_1$ & $xf_1$ & $xG$ & $0$ & even & even & $e^{(1)}$ \\
$xh_1^\perp$ & $xh_1^\perp$ & $xH^\perp$ & $2$ & even & even & $e$ \\ \hline
$xg_1$ & $xg_{1L}$ & $-x\Delta G_L$ & $0$ & even & odd & $0$ \\
$xh_{1L}^\perp$ & $xh_{1L}^\perp$ & $-x\Delta H_L^\perp$ & $2$ & odd & even & $0$ \\
$xf_{1T}^\perp$ & $xf_{1T}^\perp$ & $-xG_T$ & $1$ & odd & odd & $e_T^{(1)}$ \\
$xg_{1T}$ & $xg_{1T}$ & $-x\Delta G_T$ & $1$ & even & even & $0$ \\
$xh_{1}$ & $xh_{1T} + xh_{1T}^{\perp(1)}$ & $-x\Delta H_T$ & $1$ & odd & odd & $e_T^{(1)}$ \\
$xh_{1T}^\perp$ & $xh_{1T}^\perp$ & $-x\Delta H_T^\perp$ & $3$ & odd & odd & $-e_T$ \\ \hline
$xf_{1LL}$ &  &  & $0$ & even & even & $e_{LL}^{(1)}$ \\
$xh_{1LL}^\perp$ & &  & $2$ & even & even & $e_{LL}$ \\
$xf_{1LT}$ & &  & $1$ & even & odd & $e_{LT}^{(1)}/2$ \\
$xg_{1LT}$ & &  & $1$ & odd & even & $0$ \\
$xh_{1LT}$ & &  & $1$ & even & odd & $e_{LT}^{(1)}/2$ \\
$xh_{1LT}^\perp$ & &  & $3$ & even & odd & $-e_{LT}$ \\
$xf_{1TT}$ & &  & $2$ & even & even & $e_{TT}^{(1)}/3$ \\
$xg_{1TT}$ & &  & $2$ & odd & odd & $0$ \\
$xh_{1TT}$ & &  & $0$ & even & even & $e_{TT}^{(2)}$ \\
$xh_{1TT}^\perp$ & &  & $2$ & even & even & $-2\,e_{TT}^{(1)}/3$ \\
$xh_{1TT}^{\perp\perp}$ & &  & $4$ & even & even & $e_{TT}$ \\
\hline
\end{tabular}}
\caption{An overview of the leading twist gluon TMDs for unpolarized, vector polarized, and tensor polarized hadrons. In the second and third column, the names of the functions in this paper are compared to the ones in refs.~\protect{\cite{Mulders:2000sh,Meissner:2007rx}}. In the fourth column we list the rank of the function. Furthermore, we list the properties (even/odd) under time reversal ($T$) and charge conjugation ($C$), see appendix~\ref{a:constraints}. In the last column it is indicated to which $e$-type function the TMD reduces in the limit $x \to 0$. As a shorthand, we use the moment notation $f_{\ldots}^{(n)}(x,\bm{k}_\st^2) \equiv [\bm{k}_\st^2/(2M^2)]^n \,f_{\ldots}(x,\bm{k}_\st^2)$.}
\label{t:summary}
\end{center}
\end{table}

We end with a brief discussion of the experimental possibilities to study the gluon TMDs. The unpolarized and vector polarized gluon TMDs could be investigated in processes at RHIC, at the LHC, possibly at a future polarized fixed-target experiment at the LHC called AFTER@LHC~\cite{Brodsky:2012vg}, and at an EIC~\cite{Boer:2011fh}. For instance, the unpolarized gluon TMDs could be studied at the LHC and at AFTER@LHC in (pseudo)scalar $C$-even heavy quarkonium production, such as $\chi_{c,b}$ and $\eta_{b,c}$ in the color-singlet configuration~\cite{Boer:2012bt,Signori:2016jwo}. Another option is to consider pseudovector quarkonium such as $J/\Psi$ and $\Upsilon$, which is predominantly in the color-singlet configuration when produced in gluon fusion together with an additional isolated photon in the final state~\cite{Dunnen:2014eta}. The latter is also useful to investigate the QCD evolution of the gluon distributions. 

The linearly polarized gluon TMDs could be studied by measuring $\text{cos}(2\phi)$ modulations in processes such as dijet or heavy quark pair production in electron-proton or electron-nucleus collisions~\cite{Pisano:2013cya,Dumitru:2015gaa} and 
in virtual photon-jet pair production in $pp$ or $pA$ collisions~\cite{Metz:2011wb}. They can also be accessed through heavy quarkonium production in (un)polarized $pp$ collisions~\cite{Boer:2012bt,Signori:2016jwo} in association with other gluon TMDs.

The most promising processes that directly give access to the gluon Sivers effect are $p^\uparrow p \to \gamma \, \text{jet} \, X$ at RHIC and AFTER@LHC~\cite{Bacchetta:2007sz}, $p^\uparrow p \to J/\Psi \, \gamma\, X$ or $p^\uparrow p \to J/\Psi \, J/\Psi\, X$ at AFTER@LHC~\cite{Lansberg:2015lva}, and $e p^\uparrow \to e'\, c\bar{c}\, X$ at an EIC~\cite{Boer:2015vso}. Production of color-singlet heavy quarkonium states~\cite{Signori:2016jwo} and of photon pairs from polarized proton collisions~\cite{Qiu:2011ai} are also valid possibilities.

For some of these processes TMD factorization has not been proven yet, neither for general $x$ nor for small $x$. In order to experimentally probe the functions that remain in the small-$x$ limit, additional processes such as DIS, Drell-Yan, semi-inclusive DIS, or $p\, A \to h\, X$ offer possibilities. For a discussion and a more detailed list of relevant processes see refs.~\cite{Dominguez:2011saa,Boer:2015pni,Angeles-Martinez:2015sea}. 

The study of tensor polarized gluon TMDs would be possible at the experiments proposed to investigate polarized deuterons, e.g.\ at the EIC option put forward at Jefferson Lab (JLEIC)~\cite{Boer:2011fh,Abeyratne:2012ah,Abeyratne:2015pma}, or at COMPASS~\cite{Ball:2006zz}, although there the region of small $x$ is very limited.

\newpage
\acknowledgments
We would like to acknowledge useful discussions with Paul Hoyer and Jian Zhou. This research is part of the research program of the ``Stichting voor Fundamenteel Onderzoek der Materie (FOM)", which is financially supported by the ``Nederlandse Organisatie voor Wetenschappelijk Onderzoek (NWO)" as well as the EU FP7 ``Ideas" programme QWORK (contract no. 320389). YZ is financially supported by the China Scholarship Council.

\appendix
\section{Constraints on correlators} \label{a:constraints}
The gluon-gluon and Wilson loop correlators in this paper are constrained by hermiticity and parity ($P$). In the parametrizations hermiticity ensures that the functions are real and parity conservation only allows for $P$-even terms. Time reversal ($T$) transformations relate correlators with time-reversed gauge link structures, e.g.\ time reversal interchanges the staple-like $[+]$ and $[-]$ gauge links. Hence, time reversal invariance is not used as a constraint in the parametrizations of the correlators. For the gluon-gluon correlator the constraints are as follows:
\begin{eqnarray}
    \mbox{Hermiticity:} &\quad& \Gamma^{[U,U^\prime]\,\rho\sigma;\mu\nu\ast}(k,P,S,T,n) = \Gamma^{[U,U^\prime]\,\mu\nu;\rho\sigma}(k,P,S,T,n) , \\
    \mbox{Parity:} &\quad& \Gamma^{[U,U^\prime]\,\mu\nu;\rho\sigma}(k,P,S,T,n) = \Gamma_{\mu\nu;\rho\sigma}^{[U,U^\prime]}(\bar k,\bar P,-\bar S, \bar T,\bar n) , \\
    \mbox{Time reversal:} &\quad& \Gamma^{[U,U^\prime]\,\mu\nu;\rho\sigma\ast}(k,P,S,T,n) = \Gamma_{\mu\nu;\rho\sigma}^{[U^T,U^{\prime T}]}(\bar k,\bar P,\bar S, \bar T,\bar n) , \label{e:treversal}
\end{eqnarray}
where we have introduced the notation $\bar{a}^\mu \equiv \delta^{\mu\nu} a_\nu$ and $\bar{b}^{\mu\nu} \equiv \delta^{\mu\rho} \delta^{\nu\sigma} b_{\rho\sigma}$. Concerning the gauge links, these constraints are based on the properties $U_{[0,\xi]}^\dagger = U_{[\xi,0]}$, $U^P_{[0,\xi]} = U_{[\bar 0,\bar\xi]}$, and $U^T_{[0,\xi]} = U_{[-\bar 0,-\bar\xi]}$. By omitting the gauge links from the gluon-gluon correlator, the dependence on $n$ is no longer present. Furthermore, the gluon-gluon correlator is antisymmetric in both the pair of indices $\mu,\nu$ and $\rho,\sigma$.

For the Wilson loop correlator the constraints read:
\begin{eqnarray}
    \mbox{Hermiticity:} &\quad& \Gamma_0^{[U,U^\prime]\ast}(k;P,S,T,n) = \Gamma_0^{[U,U^\prime]}(k;P,S,T,n) , \\
    \mbox{Parity:} &\quad& \Gamma_0^{[U,U^\prime]}(k;P,S,T,n) = \Gamma_0^{[U,U^\prime]}(\bar k;\bar P,-\bar S,\bar T,\bar n) , \\
    \mbox{Time reversal:} &\quad& \Gamma_0^{[U,U^\prime]\ast}(k;P,S,T,n) = \Gamma_0^{[U^T,U^{\prime T}]}(\bar k;\bar P,\bar S,\bar T,\bar n) .
\end{eqnarray}
By simply omitting the spin vector and/or tensor from the expressions above, we obtain the constraints that apply to unpolarized (omitting $S$ and $T$), vector polarized (omitting $T$), and tensor polarized (omitting $S$) hadrons. 

The effect of charge conjugation ($C$) symmetry for gluons is found by writing down the conjugate correlator involving the conjugate field $A_\mu^{c} = -A_\mu^\dagger = -A_\mu$, such that $F^{\mu \nu \,c} = - F^{\mu \nu}$ and $U_{[0,\xi]}^c = U_{[0,\xi]}^\dagger$. As for the quark case, charge conjugation does not really give a constraint (hermiticity has already been used), but it enables us to connect partons at negative $x$ (and $\bm k_\st)$ to antipartons at positive $x$ (and $\bm k_\st)$, which is relevant for sum rules and for the relation to antiprotons. For gluons the $C$-behavior only becomes important due to the gauge link structure, in particular for the situation with two different gauge links. The correlator for the charge conjugated fields becomes
\begin{equation}
    \Gamma^{c[U,U^\prime]\mu\nu;\rho\sigma}(x,\bm k_\st) = \Gamma^{[U^{\dagger},U^{\prime\dagger}]\mu\nu;\rho\sigma}(x,\bm k_\st) , 
\end{equation}
which after rewriting and using the hermiticity constraint can also be related to the correlator at negative $x$ (and $\bm k_\st$),
\begin{equation}
    \Gamma^{c[U,U^\prime]\mu\nu;\rho\sigma}(x,\bm k_\st) = 
    \Gamma^{[U^{\prime\dagger},U^{\dagger}]\mu\nu;\rho\sigma\ast}(-x,-\bm k_\st) =
    \Gamma^{[U^{\prime\dagger},U^\dagger]\rho\sigma;\mu\nu}(-x,-\bm k_\st).
\end{equation}

For the TMDs, depending on rank and symmetry in Lorentz indices, the relations become either
\begin{equation}
    f_1^{[U,U^{\prime}]}(-x,\bm k_\st^2) = - f_1^{[U^{\prime},U]}(x,\bm k_\st^2),
\end{equation}
or  
\begin{equation}
    g_1^{[U,U^{\prime}]}(-x,\bm k_\st^2) = + g_1^{[U^\prime,U]}(x,\bm k_\st^2),
\end{equation}
referred to as $C$-even and $C$-odd respectively. The TMDs $f_1$, $h_1^\perp$, $h_{1L}^\perp$, $g_{1T}$, $f_{1LL}$, $h_{1LL}^\perp$, $g_{1LT}$, $f_{1TT}$, $h_{1TT}$ $h_{1TT}^\perp$, and $h_{1TT}^{\perp\perp}$ are $C$-even, whereas $g_1$, $f_{1T}^\perp$, $h_1$, $h_{1T}^\perp$, $f_{1LT}$, $h_{1LT}$, $h_{1LT}^\perp$, and $g_{1TT}$ are $C$-odd.

The $C$-property is of special interest for the Wilson loop correlator or in general for correlators containing a loop ${\rm Tr}_c(U_{[0,\xi]}\,U_{[\xi,0]}^{\prime})$ = ${\rm Tr}_c(U_{[0,\xi]}\,U_{[0,\xi]}^{\prime\dagger})$. One has the additional property $\Gamma_0^{[U^\dagger,U^{\prime\dagger}]}(\bm k_\st)$ = $\Gamma_0^{[U^\prime,U]}(\bm k_\st)$, thus one finds
\begin{equation}
    \Gamma_0^{c[U,U^\prime]}(\bm k_\st) = 
    \Gamma_0^{[U^\dagger,U^{\prime\dagger}]}(\bm k_\st) =
    \Gamma_0^{[U^\prime,U]}(\bm k_\st) = 
    \Gamma_0^{[U^{\prime\dagger},U^{\dagger}]}(-\bm k_\st) =
    \Gamma_0^{[U,U^\prime]}(-\bm k_\st).
\end{equation}
Hence in the Wilson loop correlator the $C$-even and $C$-odd functions can be directly identified with the even and odd rank functions. For the TMDs in the correlator $\Gamma^{[\Box]}(\bm k_\st)$ the functions $e$, $e_{LL}$, and $e_{TT}$ are $C$-even and the functions $e_T$ and $e_{LT}$ are $C$-odd. The $C$-even and $C$-odd functions are also the ones that would appear in the correlators $\big( \Gamma_0^{[\Box]}(\bm k_\st) \pm \Gamma_0^{[\Box^\dagger]}(\bm k_\st) \big) /2$ respectively. The $C$-behavior of the TMDs in the gluon-gluon and Wilson loop correlators is consistent with the small-$x$ matching in section~\ref{s:smallx}.

\section{Definitions of TMDs} \label{a:distributions}
In this appendix the definitions of the various TMDs are given in terms of the coefficients $A_i$ and $B_i$ that have been introduced in the parametrizations at the level of the unintegrated correlators. 

\subsection{The gluon-gluon correlator}
Let us denote by $\Gamma(k)$ the gluon-gluon correlator for any type of polarization,\footnote{Lorentz indices are omitted for simplicity, since they are not relevant here.} then the light-front correlator is defined as\begin{equation}
    \Gamma(x,\bm{k}_\st) \equiv \int dk\cd P \;\Gamma(k) = \frac{M^2}{2} \int [d\sigma d\tau] \,\Gamma(k) ,
\end{equation}
where we have introduced the shorthand notation
\begin{equation}
    [d\sigma d\tau] \equiv d\sigma d\tau \,\delta \left( \tau - x\sigma + x^2 + \frac{\bm{k}_\st^2}{M^2} \right) ,
\end{equation}
with the dimensionless invariants $\sigma$ and $\tau$ given by
\begin{equation}
    \sigma \equiv \frac{2k \cd P}{M^2} , \quad \tau \equiv \frac{k^2}{M^2} ,
\end{equation} 
spanning regions in remnant mass $M_R^2 \equiv (P-k)^2$ and in the partonic virtuality $k^2$. For both of these, the main contribution comes from small (hadronic) values (i.e.\ $\sigma$ and $\tau$ of order one).

The (leading twist) TMDs that occur in the parametrization of the gluon-gluon correlator for the various types of polarization in eqs.~\eqref{e:gamma_up_tmds}, \eqref{e:gamma_vp_tmds}, and \eqref{e:gamma_tp_tmds}, are related to the coefficients $A_i$ as follows:
\begin{align}
    x f_1(x,\bm{k}_\st^2) &\equiv M^2 \int [d\sigma d\tau] \left( A_2 + 2x A_4 + x^2 A_3 + \frac{\bm{k}_\st^2}{2M^2} \,A_6 \right) , \\
    x h_1^\perp(x,\bm{k}_\st^2) &\equiv M^2 \int [d\sigma d\tau] \,A_6 , \\
    x g_{1}(x,\bm{k}_\st^2) &\equiv 2M^2 \int [d\sigma d\tau] \left\{ \vphantom{\frac{\bm{k}_\st^2}{M^2}} A_8 + A_9 + x \left( A_{10} + A_{11} \right) + \left( \frac{\sigma}{2} - x \right) \left[ A_{12} + x \left( A_{14} + A_{15} \right) \right.\right. \nn \\
    &\qquad\qquad\qquad\qquad\! \left.\left. + \,x^2 A_{13} \right] + \frac{\bm{k}_\st^2}{2M^2} \left[ A_{19} + A_{23} + \left( \frac{\sigma}{2} - x \right) A_{25} \right] \right\} , \\
    x h_{1L}^\perp(x,\bm{k}_\st^2) &\equiv -2M^2 \int [d\sigma d\tau] \left[ A_{18} + A_{22} + \left( \frac{\sigma}{2} - x \right) A_{24} \right] , \\
    x f_{1T}^\perp(x,\bm{k}_\st^2) &\equiv M^2 \int [d\sigma d\tau] \left[ A_{16} - A_{20} + x \left( A_{18} - A_{22} \right) \right] , \\
    x g_{1T}(x,\bm{k}_\st^2) &\equiv -M^2 \int [d\sigma d\tau] \left[ 2A_{12} + A_{17} + A_{21} + 2x \left( A_{14} + A_{15} \right) + x \left( A_{19} + A_{23} \right) \vphantom{\frac{\bm{k}_\st^2}{M^2}} \right. \nn \\
    &\qquad\qquad\qquad\qquad \left. + \,2x^2 A_{13} + \frac{\bm{k}_\st^2}{M^2} \,A_{25} \right] , \\
    x h_{1}(x,\bm{k}_\st^2) &\equiv 2M^2 \int [d\sigma d\tau] \left[ A_{16} + A_{20} + x \left( A_{18} + A_{22} \right) + \frac{\bm{k}_\st^2}{2M^2} \,A_{24} \right] , \\
    x h_{1T}^\perp(x,\bm{k}_\st^2) &\equiv 2M^2 \int [d\sigma d\tau] \,A_{24} , \\
    x f_{1LL}(x,\bm{k}_\st^2) &\equiv \frac{M^2}{3} \int [d\sigma d\tau] \left\{ A_{27} - 2A_{34} + 2x A_{28} + x^2 A_{26} + 2 (\sigma-2x) \left( A_{37} + x A_{35} \right) \vphantom{\frac{\bm{k}_\st^2}{M^2}} \right. \nn \\
    &\qquad\qquad\qquad\quad\;\;\, \left. + \,\frac{(\sigma-2x)^2}{2} \left( A_{40} + 2x A_{42} + x^2 A_{41} \right) - \frac{\bm{k}_\st^2}{M^2} \left[ A_{26} - A_{32} \vphantom{\frac{\bm{k}_\st^2}{M^2}} \right.\right. \nn \\
    &\qquad\qquad\qquad\quad\;\;\, \left.\left. + \,A_{40} + 2x A_{42} + x^2 A_{41} + (\sigma-3x) A_{30} \right.\right. \nn \\
    &\qquad\qquad\qquad\quad\;\;\, \left.\left. - \left( \frac{(\sigma-2x)^2}{4} - \frac{\bm{k}_\st^2}{2M^2} \right) A_{44} \right] \right\} , \\
    x h_{1LL}^{\perp}(x,\bm{k}_\st^2) &\equiv - \frac{2M^2}{3} \int [d\sigma d\tau] \left[ A_{26} - A_{32} + (\sigma-3x) A_{30} \vphantom{\left( \frac{(\sigma-2x)^2}{4} - \frac{\bm{k}_\st^2}{2M^2} \right)} \right. \nn \\
    &\qquad\qquad\qquad\qquad\quad\! \left. - \,\left( \frac{(\sigma-2x)^2}{4} - \frac{\bm{k}_\st^2}{2M^2} \right) A_{44} \right] , \\
    x f_{1LT}(x,\bm{k}_\st^2) &\equiv -M^2 \int [d\sigma d\tau] \left\{ A_{37} + x A_{35} + \left( \frac{\sigma}{2} - x \right) \left( A_{40} + 2x A_{42} + x^2 A_{41} \right) \vphantom{\frac{\bm{k}_\st^2}{2M^2}} \right. \nn \\
    &\qquad\qquad\qquad\qquad\; \left. - \,\frac{\bm{k}_\st^2}{4M^2} \left[ A_{30} + \left( x - \frac{\sigma}{2} \right) A_{44} \right] \right\} , \\
    x g_{1LT}(x,\bm{k}_\st^2) &\equiv -\frac{M^2}{2} \int [d\sigma d\tau] \left[ A_{29} + \left( x - \frac{\sigma}{2} \right) \left( A_{33} + xA_{31} \right) \right] , \\
    x h_{1LT}(x,\bm{k}_\st^2) &\equiv \frac{M^2}{2} \int [d\sigma d\tau] \left\{ A_{28} + x A_{26} + \left( \frac{\sigma}{2} - x \right) \left( A_{32} + x A_{30} \right) \vphantom{\frac{\bm{k}_\st^2}{2M^2}} \right. \nn \\
    &\qquad\qquad\qquad\quad\;\;\, \left. + \,\frac{\bm{k}_\st^2}{2M^2} \left[ A_{30} + \left( x - \frac{\sigma}{2} \right) A_{44} \right] \right\} , \\
    x h_{1LT}^{\perp}(x,\bm{k}_\st^2) &\equiv -M^2 \int [d\sigma d\tau] \left[ A_{30} + \left( x - \frac{\sigma}{2} \right) A_{44} \right] , \\
    x f_{1TT}(x,\bm{k}_\st^2) &\equiv \frac{M^2}{2} \int [d\sigma d\tau] \left( A_{40} + 2x A_{42} + x^2 A_{41} + \frac{\bm{k}_\st^2}{6M^2} A_{44} \right) , \\
    x g_{1TT}(x,\bm{k}_\st^2) &\equiv \frac{M^2}{2} \int [d\sigma d\tau] \left( A_{33} + x A_{31} \right) , \\
    x h_{1TT}(x,\bm{k}_\st^2) &\equiv -\frac{M^2}{2} \int [d\sigma d\tau] \left[ A_{27} + 2x A_{28} + x^2 A_{26} + \frac{\bm{k}_\st^2}{M^2} \left( A_{32} + x A_{30} \right) \right. \nn \\
    &\qquad\qquad\qquad\qquad\; \left. - \,\frac{\bm{k}_\st^4}{4M^4} A_{44} \right] , \\
    x h_{1TT}^{\perp}(x,\bm{k}_\st^2) &\equiv \frac{M^2}{2} \int [d\sigma d\tau] \left( A_{32} + x A_{30} - \frac{\bm{k}_\st^2}{3M^2} A_{44} \right) , \\
    x h_{1TT}^{\perp\perp}(x,\bm{k}_\st^2) &\equiv \frac{M^2}{2} \int [d\sigma d\tau] \,A_{44} .
\end{align}

\subsection{The Wilson loop correlator}
For the Wilson loop correlator, translation invariance in the $\xi\cd P$ direction forces $k\cd n = x$ to be zero and the integration over $x$ is actually naturally the first to be done, even before the integration over $k\cd P$. The remaining dependence is on the invariant $k^2$, which for vanishing $x$ is just $k^2 = k_\st^2 = -\bm k_\st^2$. The TMDs in the parametrization of the Wilson loop correlator for the various types of polarization in eqs.~\eqref{e:gamma0_up_tmds}, \eqref{e:gamma0_vp_tmds}, and \eqref{e:gamma0_tp_tmds} depend on $t = k^2$ and are related to the coefficients $B_i$ as follows:
\begin{align}
    e(\bm{k}_\st^2) &\equiv \frac{M^2}{2\pi L} \int dx\,d\sigma \,B_1 , \\
    e_T(\bm{k}_\st^2) &\equiv \frac{M^2}{2\pi L} \int dx\,d\sigma \,B_2 , \\
    e_{LL}(\bm{k}_\st^2) &\equiv -\frac{M^2}{4\pi L} \int dx\,d\sigma \left[ 2B_4 + (\sigma-2x) B_5 + \left( \frac{(\sigma-2x)^2}{2} - \frac{\bm{k}_\st^2}{M^2} \right) B_3 \right] , \\
    e_{LT}(\bm{k}_\st^2) &\equiv -\frac{M^2}{4\pi L} \int dx\,d\sigma \left[ B_5 + (\sigma-2x) B_3 \right] , \\
    e_{TT}(\bm{k}_\st^2) &\equiv \frac{M^2}{4\pi L} \int dx\,d\sigma \,B_3 .
\end{align}

\section{Symmetric traceless tensors and TMDs in $\bm{b_\st}$-space} \label{a:bt}
\subsection{Symmetric traceless tensors}  \label{ss:STT}
In this appendix we list the completely symmetric and traceless tensors $k_\st^{i_1 \ldots i_n}$ that are built from the partonic momentum $k_\st$. Up to rank $n=4$, these are given by
\begin{align}
    k_\st^{ij} \equiv& \;k_\st^i k_\st^j + \frac{1}{2} \bm{k}_\st^2 g_\st^{ij} , \\
    k_\st^{ijk} \equiv& \;k_\st^i k_\st^j k_\st^k + \frac{1}{4} \bm{k}_\st^2 \left(
    g_\st^{ij} k_\st^k + g_\st^{ik} k_\st^j + g_\st^{jk} k_\st^i \right) , 
    \label{kT3}\\
    k_\st^{ijkl} \equiv& \;k_\st^i k_\st^j k_\st^k k_\st^l + \frac{1}{6} \bm{k}_\st^2 \left( g_\st^{ij} k_\st^{kl} + g_\st^{ik} k_\st^{jl} + g_\st^{il} k_\st^{jk} + g_\st^{jk} k_\st^{il} + g_\st^{jl} k_\st^{ik} + g_\st^{kl} k_\st^{ij} \right) \nn \\
    & - \frac{1}{8} \bm{k}_\st^4 \left( g_\st^{ij} g_\st^{kl} + g_\st^{ik} g_\st^{jl} + g_\st^{il} g_\st^{jk} \right) ,
\end{align}
satisfying
\begin{equation}
    {g_{\st}}_{ij} k_\st^{ij} = {g_{\st}}_{ij} k_\st^{ijk} = {g_{\st}}_{ij} k_\st^{ijkl} = 0 .
\end{equation}
Products of $k_\st$ can be decomposed into symmetric traceless tensors as follows:
\begin{align}
    k_\st^i k_\st^\alpha =& \;k_\st^{i\alpha} - \frac{1}{2} \bm{k}_\st^2 g_\st^{i\alpha} , \\
    k_\st^i k_\st^{\alpha\beta} =& \;k_\st^{i\alpha\beta} - \frac{1}{4} \bm{k}_\st^2 \left( g_\st^{i\alpha} k_\st^{\beta} + g_\st^{i\beta} k_\st^{\alpha} - g_\st^{\alpha\beta} k_\st^{i} \right) , \\
    k_\st^{ij} k_\st^{\alpha\beta} =& \;k_\st^{ij\alpha\beta} - \frac{1}{6} \bm{k}_\st^2 \left( g_\st^{i\alpha} k_\st^{j\beta} + g_\st^{i\beta} k_\st^{j\alpha} + g_\st^{j\alpha} k_\st^{i\beta} + g_\st^{j\beta} k_\st^{i\alpha} - 2 g_\st^{ij} k_\st^{\alpha\beta} - 2 g_\st^{\alpha\beta} k_\st^{ij} \right) \nn \\
    & + \frac{1}{8} \bm{k}_\st^4 \left( g_\st^{i\alpha} g_\st^{j\beta} + g_\st^{i\beta} g_\st^{j\alpha} - g_\st^{ij} g_\st^{\alpha\beta} \right) .
\end{align}
The symmetric traceless tensor $k_\st^{i_1 \ldots i_n}$ of rank $n \geq 1$ only has two independent components. This allows for a decomposition in polar coordinates:
\begin{equation}
    k_\st^{i_1 \ldots i_n} \to \frac{|\bm{k}_\st|^n}{2^{n-1}} \,e^{\pm in\varphi} ,
    \label{e:STT} 
\end{equation}
in terms of two real numbers $|\bm{k}_\st|$ and $\varphi$.

\subsection{TMDs in $b_\st$-space} \label{ss:Fourier_transforms}
Mathematically, TMD factorization decomposes a cross section as a product of functions in $\bm{b}_\st$-space, where $\bm{b}_\st$ is Fourier conjugate to the partonic transverse momentum.
As a byproduct, TMD evolution is multiplicative in $\bm{b}_\st$-space. For these reasons, it is useful to consider the light-front correlators as a function of $\bm{b}_\st$. We define correlators and functions in $\bm{b}_\st$-space as Fourier transforms of the ones in $\bm{k}_\st$-space:
\begin{align}
    \tilde{\Gamma}^{ij}(x,\bm{b}_\st) &\equiv \int d^2\bm{k}_\st \,e^{i \bm{k}_\st \cd \bm{b}_\st} \,\Gamma^{ij}(x,\bm{k}_\st)\ , \label{e:FT_corr} \\
    \tilde{f}(x,\bm{b}_\st^2) &\equiv \int d^2\bm{k}_\st \,e^{i \bm{k}_\st \cd \bm{b}_\st} f(x,\bm{k}_\st^2) \ . \label{e:FT_functions} 
\end{align}

Computing directly eq.~\eqref{e:FT_corr}, we can see that the functions entering the parametrizations of $\tilde{\Gamma}^{ij}(x,\bm{b}_\st)$ are not the ones in eq.~\eqref{e:FT_functions}, but their $n$-th derivatives with respect to $\bm{b}_\st^2$, $n$ being the rank of the function in $\bm{k}_\st$-space:
\begin{align}
    \tilde{f}^{(n)}(x,\bm{b}_\st^2) &\equiv n! \left( -\frac{2}{M^2} \frac{\partial}{\partial \bm{b}_\st^2} \right)^n \tilde{f}(x,\bm{b}_\st^2) \nn \\
    &= \frac{2\pi n!}{M^{2n}} \int_0^{\infty} d|\bm{k}_\st| \,|\bm{k}_\st| \left( \frac{|\bm{k}_\st|}{|\bm{b}_\st|} \right)^n J_n(|\bm{k}_\st| |\bm{b}_\st|) \,f(x,\bm{k}_\st^2) ,
    \label{e:nth_derivative_bT2}
\end{align}
where $J_n(z)$ is the Bessel function of the first kind of order $n$, which is defined as
\begin{equation}
    J_n(z) = \frac{1}{2\pi i^n} \int_0^{2\pi} d\varphi \,e^{in\varphi} e^{iz \cos \varphi} .
    \label{e:def_Jn}
\end{equation}
In eq. \eqref{e:nth_derivative_bT2} we also used the relation
\begin{equation}
    \left( \frac{1}{z} \frac{d}{dz} \right)^k \left( z^{-\nu} J_\nu(z) \right) = (-1)^k z^{-\nu-k} J_{\nu+k}(z) ,
    \label{e:der_Bessel}
\end{equation}
considering $\nu=0$, $k=n$, and $z=|\bm{k}_\st| |\bm{b}_\st|$ with $|\bm{k}_\st|$ fixed. The factor $M^{-2n}$ renders the derivative operator dimensionless and the $n!$ is added to match the conventions in ref. \cite{Boer:2011xd}. 

From eq. \eqref{e:nth_derivative_bT2} it follows that for definite rank TMDs there is a one-to-one correspondence between the functions in $\bm{b}_\st$-space and in $\bm{k}_\st$-space. In the following subsections we provide the gluon-gluon and Wilson loop correlators in $\bm{b}_\st$-space.

\subsubsection{The gluon-gluon correlator} \label{ss:gg_corr_bT}
The light-front gluon-gluon correlator for a spin-$1$ hadron is given in $\bm{b}_\st$-space by
\begin{align}
    \tilde{\Gamma}^{ij}(x,\bm{b}_\st) =& \;\tilde{\Gamma}_U^{ij}(x,\bm{b}_\st) + \tilde{\Gamma}_L^{ij}(x,\bm{b}_\st) + \tilde{\Gamma}_T^{ij}(x,\bm{b}_\st) \nn \\
    & + \tilde{\Gamma}_{LL}^{ij}(x,\bm{b}_\st) + \tilde{\Gamma}_{LT}^{ij}(x,\bm{b}_\st) + \tilde{\Gamma}_{TT}^{ij}(x,\bm{b}_\st) ,
\end{align}
where
\begin{align}
    \tilde{\Gamma}_U^{ij}(x,\bm{b}_\st) &= \frac{x}{2} \left[ - \,g_\st^{ij} \,\tilde{f}_1(x,\bm{b}_\st^2) - \frac{M^2 \,b_\st^{ij}}{2} \,\tilde{h}_1^{\perp (2)}(x,\bm{b}_\st^2) \right] , \\
    \tilde{\Gamma}_L^{ij}(x,\bm{b}_\st) &= \frac{x}{2} \left[ i \epsilon_\st^{ij} S_L \,\tilde{g}_1(x,\bm{b}_\st^2) - \frac{M^2 \,{\epsilon_\st^{\{i}}_\alpha \,b_\st^{j\}\alpha} S_L }{4} \,\tilde{h}_{1L}^{\perp (2)}(x,\bm{b}_\st^2) \right] , \\
    \tilde{\Gamma}_T^{ij}(x,\bm{b}_\st) &= \frac{x}{2} \left[ - \,iM \,g_\st^{ij} \epsilon_\st^{S_T b_\st} \,\tilde{f}_{1T}^{\perp (1)}(x,\bm{b}_\st^2) - M \,\epsilon_\st^{ij} \,\bm{b}_\st \cd \bm{S}_T \,\tilde{g}_{1T}^{(1)}(x,\bm{b}_\st^2) \vphantom{\frac{iM^3 \,{\epsilon_\st^{\{i}}_\alpha \,b_\st^{j\}\alpha S_\st}}{12}} \right. \nn \\
    &\qquad\quad\!\! \left. - \,\frac{iM}{4} \left( \epsilon_\st^{b_\st\{i} S_\st^{j\}} + \epsilon_\st^{S_\st\{i} b_\st^{j\}} \right) \,\tilde{h}_1^{(1)}(x,\bm{b}_\st^2) \right. \nn \\
    &\qquad\quad\!\! \left. + \,\frac{iM^3 \,{\epsilon_\st^{\{i}}_\alpha \,b_\st^{j\}\alpha S_\st}}{12} \,\tilde{h}_{1T}^{\perp (3)}(x,\bm{b}_\st^2) \right] , \\
    \tilde{\Gamma}_{LL}^{ij}(x,\bm{b}_\st) &= \frac{x}{2} \left[ - \,g_\st^{ij} S_{LL} \,\tilde{f}_{1LL}(x,\bm{b}_\st^2) -\frac{M^2 \,b_\st^{ij} S_{LL}}{2} \,\tilde{h}_{1LL}^{\perp (2)}(x,\bm{b}_\st^2) \right] , \\
    \tilde{\Gamma}_{LT}^{ij}(x,\bm{b}_\st) &= \frac{x}{2} \left[ - \,iM \,g_\st^{ij} \,\bm{b}_\st \cd \bm{S}_{LT} \,\tilde{f}_{1LT}^{(1)}(x,\bm{b}_\st^2) - M \,\epsilon_\st^{ij} \epsilon_\st^{S_{LT}b_\st} \,\tilde{g}_{1LT}^{(1)}(x,\bm{b}_\st^2) \vphantom{\frac{iM^3 \,b_\st^{ij\alpha} {S_{LT}}_\alpha}{6}} \right. \nn \\
    &\qquad\quad\!\! \left. + \,iM \,S_{LT}^{\{i} \,b_\st^{j\}} \,\tilde{h}_{1LT}^{(1)}(x,\bm{b}_\st^2) - \frac{iM^3 \,b_\st^{ij\alpha} {S_{LT}}_\alpha}{6} \,\tilde{h}_{1LT}^{\perp (3)}(x,\bm{b}_\st^2) \right] , \\
    \tilde{\Gamma}_{TT}^{ij}(x,\bm{b}_\st) &= \frac{x}{2} \left[ \frac{M^2 \,g_\st^{ij} b_\st^{\alpha\beta} {S_{TT}}_{\alpha\beta}}{2} \,\tilde{f}_{1TT}^{(2)}(x,\bm{b}_\st^2) 
    - \frac{iM^2 \,\epsilon_\st^{ij} {\epsilon^{\beta}_\st}_\gamma b_\st^{\gamma\alpha} {S_{TT}}_{\alpha\beta}}{2} \,\tilde{g}_{1TT}^{(2)}(x,\bm{b}_\st^2) \right. \nn \\
    &\qquad\quad\!\! \left. + \,S_{TT}^{ij} \,\tilde{h}_{1TT}(x,\bm{b}_\st^2) - \frac{M^2 \,{S_{TT}^{\{i}}_\alpha b_\st^{j\}\alpha}}{2} \,\tilde{h}_{1TT}^{\perp (2)}(x,\bm{b}_\st^2) \right. \nn \\
    &\qquad\quad\!\! \left. + \,\frac{M^4 \,b_\st^{ij\alpha\beta} {S_{TT}}_{\alpha\beta}}{24} \,\tilde{h}_{1TT}^{\perp\perp (4)}(x,\bm{b}_\st^2) \right] .
\end{align}

\subsubsection{The Wilson loop correlator} \label{ss:Wloop_corr_bT}
The light-front Wilson loop correlator for a spin-$1$ hadron is given in $\bm{b}_\st$-space by
\begin{align}
    \tilde{\Gamma}_0^{[U,U^\prime]}(\bm b_\st) =& 
    \;\tilde{\Gamma}^{[U,U^\prime]}_{0\hspace{0.08cm}U}(\bm b_\st) + 
    \tilde{\Gamma}^{[U,U^\prime]}_{0\hspace{0.08cm}L}(\bm b_\st) + 
    \tilde{\Gamma}^{[U,U^\prime]}_{0\hspace{0.08cm}T}(\bm b_\st) \nn \\
    &+ \tilde{\Gamma}^{[U,U^\prime]}_{0\hspace{0.08cm}LL}(\bm b_\st) + 
    \tilde{\Gamma}^{[U,U^\prime]}_{0\hspace{0.08cm}LT}(\bm b_\st) + 
    \tilde{\Gamma}^{[U,U^\prime]}_{0\hspace{0.08cm}TT}(\bm b_\st) ,
\end{align}
where
\begin{align}
    \tilde{\Gamma}^{[U,U^\prime]}_{0\hspace{0.08cm}U}(\bm b_\st) &= \frac{\pi L}{M^2} \,\tilde{e}(\bm{b}_\st^2) , \\
    \tilde{\Gamma}^{[U,U^\prime]}_{0\hspace{0.08cm}L}(\bm b_\st) &= 0 , \\
    \tilde{\Gamma}^{[U,U^\prime]}_{0\hspace{0.08cm}T}(\bm b_\st) &= \frac{i\pi L}{M} \,\epsilon_\st^{S_T b_\st} \,\tilde{e}_T^{(1)}(\bm{b}_\st^2) , \\
    \tilde{\Gamma}^{[U,U^\prime]}_{0\hspace{0.08cm}LL}(\bm b_\st) &= \frac{\pi L}{M^2} \,S_{LL} \,\tilde{e}_{LL}(\bm{b}_\st^2) , \\
    \tilde{\Gamma}^{[U,U^\prime]}_{0\hspace{0.08cm}LT}(\bm b_\st) &= \frac{i\pi L}{M} \,(\bm{b}_\st \cd \bm{S}_{LT}) \,\tilde{e}_{LT}^{(1)}(\bm{b}_\st^2) , \\
    \tilde{\Gamma}^{[U,U^\prime]}_{0\hspace{0.08cm}TT}(\bm b_\st) &= -\frac{\pi L}{2} \,b_\st^{\alpha\beta} {S_{TT}}_{\alpha\beta} \,\tilde{e}_{TT}^{(2)}(\bm{b}_\st^2) .
\end{align}

\bibliographystyle{JHEP}
\bibliography{references}

\end{document}